\documentclass{optica-article}

\journal{opticajournal} 

\articletype{Research Article}
\usepackage{booktabs}
\usepackage{caption}
\usepackage{subcaption}
\usepackage{makecell}
\usepackage{amsmath}
\usepackage{siunitx}
\allowdisplaybreaks[4]
\begin{document}

\title{Approximate model for the coupling of far-field wavefront errors and jitter in space-based gravitational wave laser interferometry}
\author{Ya-Zheng Tao,\authormark{1,2}$^\dagger$ Rui-Hong Gao,\authormark{3}$\dagger$ Hong-Bo Jin,\authormark{4,5}$^\ast$ Zhen-Xiang Hao,\authormark{1,2} Gang Jin,\authormark{1} and Yue-Liang Wu,\authormark{2,5,6}}

\address{
\authormark{1}School of Fundamental Physics and Mathematical Sciences, Hangzhou Institute for Advanced Study, UCAS, Hangzhou 310024, Zhejiang Province, China\\
\authormark{2}University of Chinese Academy of Sciences, Beijing 10049, China\\
\authormark{3}Center for Gravitational Wave Experiment, National Microgravity Laboratory, Institute of Mechanics,
Chinese Academy of Sciences, Beijing 100190, China\\
\authormark{4}National Astronomical Observatories, Chinese Academy of Sciences, Beijing 100101, China\\
\authormark{5}The International Centre for Theoretical Physics Asia-Paciﬁc, University of Chinese Academy of Sciences,Beijing 100190,China\\
\authormark{6}CAS Key Laboratory of Theoretical Physics, Institute of Theoretical Physics,Chinese Academy of Sciences, Beijing 100190,China
}
\noindent\authormark{$\dagger$} These two authors contributed equally.\\
\noindent\authormark{$\ast$} Corresponding author\\
\email{\authormark{$\dagger$}taoyazheng@ucas.ac.cn}
\email{\authormark{$\dagger$}gaoruihong@imech.ac.cn}
\email{\authormark{$\ast$}hbjin@bao.ac.cn} 


\begin{abstract*} 
Space-based gravitational wave observatories, such as LISA, Taiji, and TianQin, employ long-baseline laser interferometry, necessitating displacement measurement sensitivity at 1 pm/$\sqrt{Hz}$ level. A significant challenge in achieving this precision is the coupling noise arising from far-field wavefront errors (WFE) and laser pointing jitter. This paper presents a comprehensive noise model that incorporates three critical factors: transmitted WFE, static pointing angle, and laser beam jitter. Utilizing the Nijboer-Zernike diffraction theory, we derive an approximate expression for far-field WFE, ensuring minimal error and efficient computational performance. The approximate expression has convincing physical interpretability and reveals how various Zernike aberrations and their coupling impact far-field WFE. Furthermore, the study identifies that correcting optical axis deviations induced by $Z_3^{\pm1}$ through beam tilt exacerbates far-field WFE, underscoring the necessity for active suppression of $Z_3^{\pm1}$. The proposed model facilitates detailed system simulations of the laser link, evaluates Tilt-to-Length (TTL) noise, and offers theoretical insights for system optimization.
\end{abstract*}

\section{Introduction}\label{se:1}
Typical space-based gravitational wave detection projects, such as LISA, Taiji, and TianQin, aim to detect gravitational waves in the low-to-mid frequency band of 0.1 mHz to 1 Hz \cite{jennrich2009lisa,luo2020brief,luo2016tianqin}. These projects utilize long-baseline laser interferometry with three-satellite formations as their fundamental measurement principle, with arm lengths ranging from $10^8$ to $10^9\;\text{m}$\cite{jennrich2009lisa,luo2020brief,luo2016tianqin}. Considering this arm length and considering the strain sensitivity of gravitational waves upon reaching Earth, the resolution for displacement measurements must achieve a level of 1 pm/$\sqrt{\text{Hz}}$ \cite{jennrich2009lisa,luo2020brief,luo2016tianqin}. To meet such extreme measurement requirements, the measurement methodology itself must not only achieve this precision but also effectively eliminate or suppress a significant amount of noise that can impact measurements within the relevant frequency band.

Far-field WFE and laser pointing jitter coupling noise, as one type of Tilt-to-Length (TTL) coupling, represent an unavoidable source of displacement noise in measurements\cite{wanner2024depth}. This noise arises from the interaction between the pointing angle and the far-field WFE of the laser during inter-satellite link transmission \cite{robertson1997optics,caldwell1998optical,bender2005wavefront,sasso2018coupling}. Given that it is impractical to establish an experimental setup on Earth for beam propagation over millions of kilometers, current research on this issue primarily relies on theoretical analysis and numerical simulations.

Early research began with Robertson et al.'s estimated formula for the coupling effect of curvature WFE and beam-pointing errors\cite{robertson1997optics}. Subsequently, Waluschka conducted numerical computations of the far-field wavefront for LISA's arm length using simulation software\cite{waluschka1999lisa}. In 2005, Bender utilized the Fresnel-Kirchhoff diffraction integral to analyze this issue. He expanded the exit pupil function in a third-order power series and considered the coupling effects of low-order aberrations, such as defocus and astigmatism, coupled with pointing jitter on the far-field phase \cite{bender2005wavefront}. This work provided an overall estimation of the far-field WFE level and proposed a transmitted wavefront quality criterion of $\lambda/20$. In 2018, Sasso et al. applied Bender’s method to analyze the coupling relationships between low-order Zernike aberrations and tilt. They inspected the regions in the far field by adjusting the tip/tilt in the Tx pupil and used Monte Carlo methods to study the impact of random wavefront distortions on measurement noise \cite{sasso2018coupling}. Based on this model, Sasso et al. and Zhao Y. examined the combined effects of misalignments and aberrations of the interfering wavefronts on the phase of the heterodyne signal \cite{sasso2018coupling2,zhao2020tilt}. Additionally, Zhao et al. extended this method to higher-order aberrations and investigated the far-field optical path phase on a prototype telescope \cite{zhao2021far}. Chen et al. conducted a study on telescope optimization based on the different contributions of aberrations to coupling noise \cite{chen2022reducing}. Xiao et al. further extended this method to higher-order aberrations, focusing on how the stationary point helps suppress coupling noise \cite{xiao2023analysis}. In 2019, Vinet et al. took an alternative approach by simplifying the diffraction integral through a first-order expansion of the transmitted WFE and numerically solving the derived expression to obtain the coupling relationship between pointing jitter angles and aberrations\cite{vinet2020numerical}. Their analysis revealed varying degrees of influence from different single aberrations on coupling noise. Additionally, Kenny, Weaver, and Tao conducted numerical simulations on this issue using numerical integration, the Mode Expansion Method (MEM), and the Gaussian Beam Decomposition (GBD), respectively \cite{kenny2020beam,weaver2020analytic,tao2023estimation}.

We find that the current research has certain shortcomings. While the model by Sasso et al. simplifies calculations, it may not provide strong physical interpretability. This model does not fully clarify how individual aberrations and their coupling contribute to the far-field WFE. Furthermore, while the model equates tip/tilt with pointing, it does not account for the influence of other odd Zernike aberrations, such as coma, on pointing performance \cite{kenny2020beam}. Vinet et al.'s analytical calculations using Nijboer-Zernike theory are limited to first-order approximations and do not discuss the coupling relationships between aberrations. While numerical computations and simulations yield accurate results, they are time-consuming and hinder rapid responses in the noise analysis of the entire laser link. Therefore, we aim to overcome these shortcomings. In this paper, we refer to the analysis of diffraction propagation of a single aberration under point source conditions, as described in the Nijboer-Zernike theory\cite{born2013principles}. We analytically derive the diffraction formula for distorted Gaussian beams in the far field, obtaining a model for far-field WFE and pointing jitter noise. This model  offers convincing physical interpretability, encompassing three major factors influencing noise, meeting precision requirements, and providing rapid computational responses. Specifically, Section \ref{se:2} introduces the application of Nijboer-Zernike Theory to address the far-field propagation of distorted Gaussian beams. In Section \ref{se:3}, we present a comprehensive coupling noise model that incorporates the quality of the transmitted wavefront, the static pointing angle, and laser beam jitter, along with a derived formula for quantifying the associated noise levels. Section \ref{se:4} calculates the contributions of various aberrations to the far-field WFE, revealing how different aberrations contribute to and couple with each other. This section also discusses how aberrations affect the optical axis. In Section \ref{se:5}, we present a formula for the total far-field WFE that incorporates the first 21 Zernike aberrations. Additionally, we provide an example to illustrate how to apply this formula in the discussion of coupled noise. Finally, Section \ref{se:6} provides the conclusions and summary.

\section{Nijboer-Zernike Theory and the expression of far-field wavefront error} \label{se:2}
\begin{figure*}[htbp]
	\begin{center}
		\includegraphics[width=0.9\textwidth]{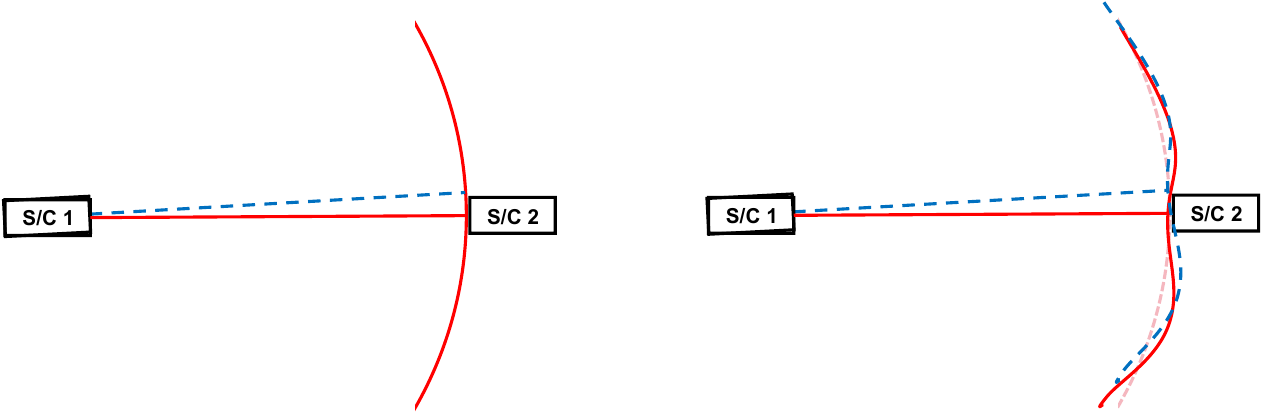}
	\end{center}
	\caption{This schematic diagram illustrates how TTL-wavefront distortion coupling occurs. On the left, the ideal transmitted beam with jitter propagates into the telescope aperture of another spacecraft. On the right, the propagation of the beam carrying WFEs is depicted. The truncated wavefront is transformed from a spherical wave (left) to a distorted wavefront (right).}
	\label{TiltSetup}
\end{figure*}

Far-field wavefront error (WFE) and coupling noise from laser pointing jitter arise from the following process: A local laser on one spacecraft (S/C1) is emitted through its local telescope, with the beam assumed to be a truncated Gaussian beam at the telescope aperture. If the wavefront is considered undistorted—meaning that the emitted wavefront is an ideal plane wave—it will approximate a spherical wavefront after propagating over millions of kilometers. At this stage, deviations in the initial propagation direction do not impart additional phase to the top-hat beam received by the telescope on the receiving spacecraft (S/C2), as illustrated on the left side of Fig. \ref{TiltSetup}. However, in practice, the emitted beam’s wavefront will be distorted due to various mechanisms, resulting in a non-ideal spherical wavefront at the receiving end, which introduces a WFE relative to the ideal spherical wavefront. Changes in the propagation direction of the beam then couple with these WFEs, causing additional angle-dependent phase fluctuations in the received flat-top beam, as shown on the right side of Fig. \ref{TiltSetup}, thereby affecting the measurement accuracy. Since phase fluctuations are directly related to the laser jitter frequency and operate within the same detection frequency band as the Tilt-to-Length (TTL) noise from the optical bench, this coupling noise is also classified as TTL noise, as both contribute to displacement noise in optical path measurements.

In accordance with this physical model, we describe the far-field WFE as the phase difference between the distorted Gaussian beam and the ideal Gaussian beam at the receiving side in the far field. Frits Zernike and Bernard Nijboer developed a diffraction theory \cite{nijboer1947diffraction}, which enables the determination of the electric field of a distorted beam in the image plane, thereby allowing the phase term to be calculated. First, we define the initial electric field at the aperture of the transmitting telescope as follows:
\begin{equation}\label{wfegeneraldefinition}
	E_a(r; 0)=E_0(r; 0)e^{i\varOmega_a},	
\end{equation}
where $E_0(r; 0)$ is the electric field of an ideal truncated circular Gaussian beam:
\begin{equation}
	E_0(r; 0)=
	\left\{
	\begin{aligned}
		e^{-r^2/w_0^2},&&\mbox{if $r\leq{r_a}$ } \\
		0,&&\mbox{if $r>{r_a}$ } 
	\end{aligned}
	\right.
\end{equation}
where $w_0$ is the waist radius of Gaussian beam and $r_a$ is the aperture radius of telescope. $\varOmega_a$ in \eqref{wfegeneraldefinition} is the transmitted WFE, which is described by a combination of a set of Zernike polynomials:
	\begin{equation}\label{AberrationDefinition}
		\varOmega_a(\rho,\theta)=\sum_{n=0}^{N}\sum_{m=-n}^{n} a^m_n\,Z^m_n(\rho,\theta),
	\end{equation}
where $a^m_n$ represents coefficients and $Z^m_n$ are Zernike Polynomals written with OSA/ANSI indexing. $\rho=r/r_a$ is restricted to unit disk (0$\leq\rho\leq$1), and $\theta$ is the azimuth. $n-m\geq0$ and is even. Zernike Polynomals is a circle polynomials:
 \begin{equation}
	Z^m_n(\rho,\theta)=
	\left\{
	\begin{aligned}
		R^m_n(\rho)\cos\left(m\theta\right),&&\mbox{if $m\geq0$ } \\
		R^{-m}_n(\rho)\sin\left(-m\theta\right),&&\mbox{if $m<0$ }
	\end{aligned}
	\right.
\end{equation}
$R^m_n(\rho)$ is the radial polynomials:
\begin{equation}
	R_n^{|m|}(\rho)=(-1)^{(n-|m|)/2}\rho^{|m|}P_{(n-|m|)/2}^{(|m|,0)}(1-2\rho^2) ,
\end{equation}
where $P_{t}^{(\alpha,\beta)}$ is the Jacobi polynomial of degree t, and $R^m_n(\rho)$ satisfies the orthogonality relation:
\begin{equation}
	\int_0^1R_n^{|m|}(\rho)R_{n'}^{|m|}(\rho)\rho{d}\rho=\frac{\delta_{n,n'}R_n^{|m|}(1)}{2(n+1)}.
\end{equation}

Considering that the pointing angle of the beam is negligible compared to the propagation distance, the diffraction integral meets the conditions of the paraxial approximation. Consequently, the electric field in the far field can be represented using Fraunhofer diffraction:
\begin{equation}\label{FraunhoferAberration1}
	E(r, \psi, z)=\frac{e^{ikz}e^{\frac{ik}{2z}r^2}}{i\lambda{z}}{r_a}^2
	\int_0^1\int_0^{2\pi}e^{-\rho^2}e^{i\varOmega_a(\rho,\theta)}e^{-iv\rho\cos{(\theta-\psi)}}\rho{d}\rho{d}\theta.
\end{equation}
where $w_0=r_a$ is assumed and $v=\frac{k}{{z}}{r_a}r$.

We will discuss the integral:
\begin{equation}\label{FraunhoferAberration2}
	U(r, \psi, z)=\int_0^1\int_0^{2\pi}e^{-\rho^2}e^{i\varOmega_a(\rho,\theta)}e^{-iv\rho\cos{(\theta-\psi)}}\rho{d}\rho{d}\theta.
\end{equation}
We expand $e^{-\rho^2}$ and $e^{i\varOmega_a(\rho,\theta)}$ within \eqref{FraunhoferAberration2} separately. Firstly, for the $e^{-\rho^2}$ part, we have Bauer formula\cite{watson1922treatise}:
\begin{equation}\label{Bauer1}
	e^{-z\cos\phi}=(\frac{\pi}{2z})^{\frac{1}{2}}\sum_{l=0}^{\infty}{(-1)^l(2l+1)I_{l+\frac{1}{2}}(z)}{P_l(\cos{\phi})},	
\end{equation}
where $I_{l+\frac{1}{2}}$ is Modified Bessel functions of the first kind, and $P_l$ is Legendre polynomial. We set $\cos\phi=2{\rho}^2-1$, and use the relation $P_l(2{\rho}^2-1)=R_{2l}^0(\rho)$, it follows:
\begin{equation}\label{Bauer2}
e^{-\rho^2}=e^{-\frac{1}{2}}e^{-\frac{1}{2}(2{\rho}^2-1)}=e^{-\frac{1}{2}}{\pi}^{\frac{1}{2}}\sum_{l=0}^{\infty}{(-1)^l(2l+1)I_{l+\frac{1}{2}}(\frac{1}{2})}{R_{2l}^{0}(\rho)}.
\end{equation}
For the accuracy of the calculation results, expanding it to the second order ($l=2$) is sufficient.

Secondly, for the $e^{i\varOmega_a(\rho,\theta)}$ component, we apply a Taylor expansion. The expanded terms may include $\cos(sin)(m\theta)$, which must be considered in the angular integral. For the angular integral of each term in the expansion, we can derive the result using Product-to-Sum identities and Bessel's integrals:
\begin{equation}\label{BesselIntegrals}
\int_0^{2\pi}\cos(sin)(m\theta)e^{-iv\rho\cos{(\theta-\psi)}}{d}\theta=(-i)^{m}2\pi{cos(sin)(m\psi)}J_m(v\rho).
\end{equation}
And for the radial integral part, after simplification, we handle the following integral:
\begin{equation}\label{RadialIntegrals}
\int_0^1{R^m_n(\rho)}{{R_{2l}^{0}(\rho)}{J_m(v\rho)}}\rho{d}\rho.
\end{equation}
For $R^{m}_{n}(\rho)R_{2l}^{0}(\rho)$ we have
\begin{equation}\label{GaussE}
	R^m_n(\rho)R_{2l}^{0}(\rho)=\sum_{j=0}^{l+\frac{n-m}{2}}{b_j}{R^m_{2j+m}(\rho)},	
\end{equation}
where
\begin{gather*}\label{relation2}
	\sum_{j=0}^{l+\frac{n-m}{2}}{b_j}=1,
\end{gather*}
and each $b_j$ can be determined by Gauss Elimination. By using the relation:
\begin{equation}
	\int_0^1{R^m_n(\rho)}{J_m(v\rho)}{\rho}{d}\rho=(-1)^{\frac{n-m}{2}}\frac{J_{n+1}(v)}{v},
\end{equation}
finally we can express the integral part of \eqref{FraunhoferAberration2} as a linear combination of a series of ${J_{n}(v)}/{v}$. 

For example, by combining $Z^{-m}_{n}$ and $Z^{m}_{n}$ as ${c^m_n}{R^m_n}(\rho)\cos(m(\theta-\theta_0))$, performing a first-order expansion for $e^{i\varOmega_a(\rho,\theta)}$ part yields the following result derived from \eqref{FraunhoferAberration2}:
\begin{subequations}\label{FraunhoferAberration3}
\begin{align}
&U(r, \psi, z)=U_0(r, \psi, z)+U_1(r, \psi, z),\\
&U_0(r, \psi, z)=\sum_{l=0}^{\infty}{(2l+1)I_{l+\frac{1}{2}}(\frac{1}{2})}\frac{J_{2l+1}(v)}{v},\\
\begin{split}
&U_1(r, \psi, z)=\sum_{n=0}^{N}\sum_{m=0}^{n}\left[{c^m_n}{(-i)^{m-1}}{\cos{m(\psi-\theta_0)}}M_{nm}\right],\\
&M_{nm}=\sum_{l=0}^{\infty}{(2l+1)I_{l+\frac{1}{2}}(\frac{1}{2})}\sum_{j=0}^{l+\frac{n-m}{2}}\frac{1}{2}{b_j}{(-1)^{j+l}}\frac{J_{2j+1}(v)}{v}.
\end{split}
\end{align}
\end{subequations}
Results for higher-order expansions can also be derived using the same process. Notably, $U_0(r, \psi, z)$ represents the undistorted part of the Gaussian beam. The far-field WFE, denoted as ${\delta}{\Theta}(r, \psi, z)$, is defined as the phase difference in the far field between the distorted Gaussian beam and its undistorted counterpart, expressed as:
\begin{equation}\label{FraunhoferAberration4}
{\delta}{\Theta}(r, \psi, z)=\frac{\lambda}{2\pi}\left(\frac{Im\left\{U(r, \psi, z)\right\}}{Re\left\{U(r, \psi, z)\right\}}-\frac{Im\left\{U(0, \psi, z)\right\}}{Re\left\{U(0, \psi, z)\right\}}\right).
\end{equation}

\section{Coupling noise model} \label{se:3}
The noise level of coupling noise arising from far-field WFE and laser pointing jitter is influenced by three primary factors: transmitted WFE (or transmitted wavefront quality), static pointing angle, and laser jitter. This section discusses the correlation between these three factors and the noise level of the coupling noise using a physical model.

The far-field WFE ${\delta}{\Theta}(r, \psi, z)$, derived in the previous section, is expressed in Cartesian coordinates as ${\delta}{\Theta}(x=r\cos\psi, y=r\sin\psi, z)$.  At a fixed distance, the far-field WFE is a function of spatial coordinates within the plane ${\delta}{\Theta}(x, y, z=L)$, where $L$ represents the arm length between the two spacecraft. In the actual physical process, the laser's pointing angle varies at the transmitter while the receiver remains stationary. This can be equivalently viewed as the transmitter's pointing angle remaining constant while the receiver's telescope moves within the plane of ${\delta}{\Theta}(x, y, z=L)$. Given that the pointing angle is very small relative to the arm length, the horizontal displacement corresponding to the pointing angle can be approximated as a linear relationship:
\begin{equation}\label{RadialPlane}
	r=L \cdot a,
\end{equation}
where a is the pointing angle. Taking Taiji as an example, with an arm length of $3\;\text{Mkm}$, a change in the pointing angle of $10\;\text{nrad}$ approximately corresponds to a horizontal displacement of $30\;\text{m}$.  Considering that the telescope's aperture size is only $0.4\;\text{m}$, the difference in WFE distribution across the telescope aperture is negligible. This is also why the received wavefront is considered to be a (clipped) plane wavefront, often referred to as a top-hat wavefront. The motion of the receiver's telescope within the plane of ${\delta}{\Theta}(x, y, z=L)$ can be regarded as the motion of a point within that plane.

Next, the pointing angle can be categorized into static and dynamic pointing angles. The static pointing angle primarily results from manufacturing or assembly errors of the telescope, deformation of the mirrors due to thermal effects, or other slow-changing processes while in orbit. We assume these factors introduce a shift of $(x_0, y_0)$ on the far-field WFE plane. The dynamic pointing angles mainly refer to laser jitter, which causes instantaneous changes in the pointing angle $({\theta}_x, {\theta}_y)$. Additionally, the static pointing angle is associated with the shift on the plane as described by \eqref{RadialPlane}. The level of laser jitter is denoted as $J$, forming a circular area on the far-field WFE plane ${\delta}{\Theta}(r, \psi, z=L)$, within which the shift caused by the dynamic pointing angle is confined. Therefore, we can summarize the above model as follows:
\begin{equation}\label{WFE}
	\left\{
	\begin{aligned}	
	&{W}_E({\theta}_x,{\theta}_y)= {\delta}{\Theta}(L\cdot{\theta}_x+x_0,L\cdot{\theta}_y+y_0,z=L),\\
	& \sqrt{{{\theta}_x}^2+{{\theta}_y}^2}\leq J,		
\end{aligned}
\right.
\end{equation}
where ${W}_E({\theta}_x,{\theta}_y)$ represents the possible range of magnitudes for the far-field WFE. It also indicates the level of phase offset in the received top-hat beam. The noise level of the coupling noise can be represented by the magnitude of the gradient of the WFE within the circular area, ${W}_E({\theta}_x,{\theta}_y)$, which is:
\begin{equation}\label{NoiseLevel}
{\delta}({\theta}_x,{\theta}_y)= \Vert \nabla ({W}_E({\theta}_x,{\theta}_y)) \Vert =\sqrt{ {(\frac{\partial{{W}_E({\theta}_x,{\theta}_y)}}{\partial{{\theta}_x}})}^2+{(\frac{\partial{{W}_E({\theta}_x,{\theta}_y)}}{\partial{{\theta}_y}})}^2}.
\end{equation}
We therefore relate the noise level of the coupling noise to the transmitted WFE, static pointing angle, and laser jitter through \eqref{WFE} and \eqref{NoiseLevel}.

\section{Far field calculation} \label{se:4}
\begin{figure*}[htbp]
	\begin{center}
		\includegraphics[width=0.75\textwidth]{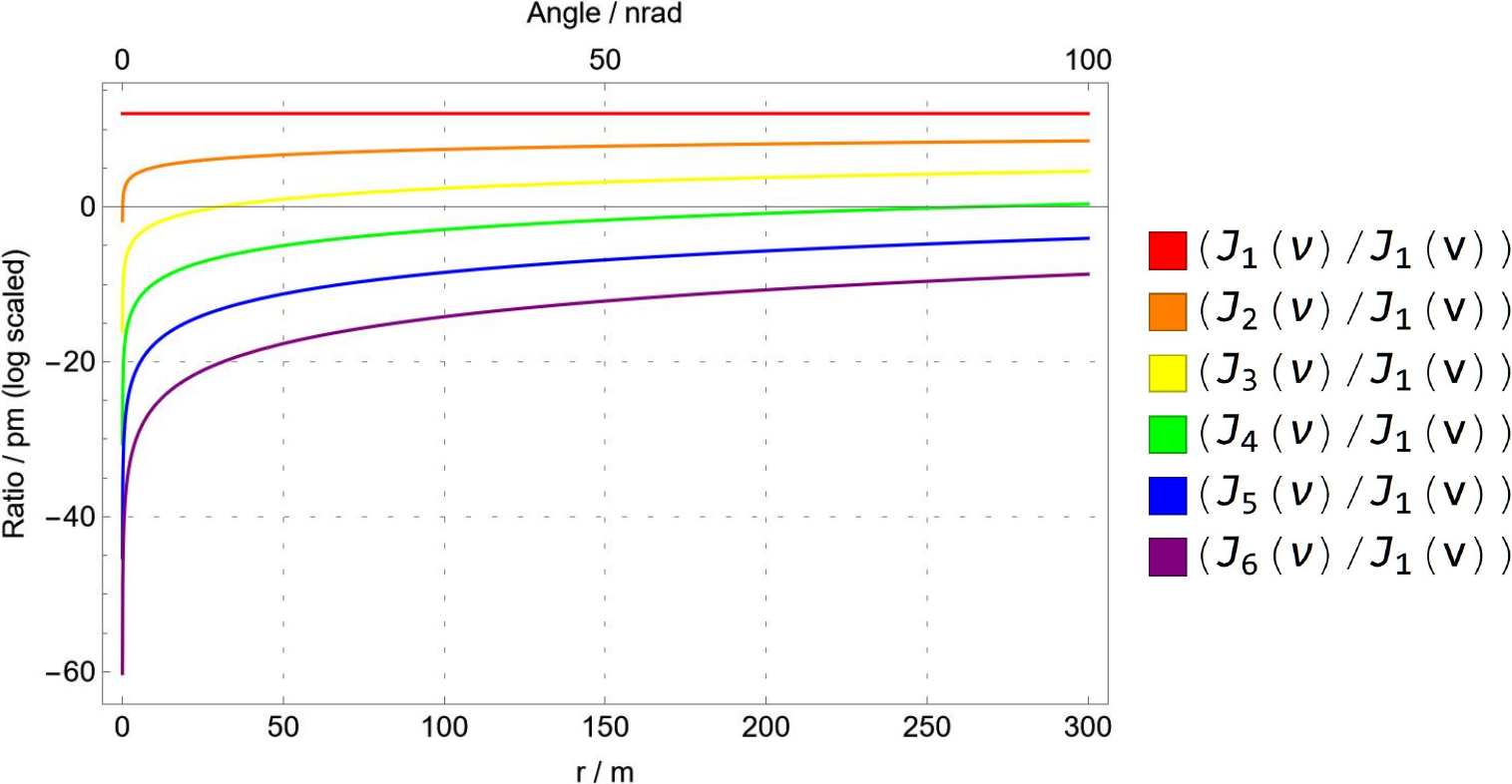}
	\end{center}
	\caption{The ratio of Bessel functions $J_m(v)$ ($m\leq6$) to $J_1(v)$. This ratio is converted to units of "$\text{pm}$" by multiplying by the coefficient factor "$\frac{\lambda}{2\pi}$", where $\lambda=1.064 \times 10^{6}\;\text{pm}$. In the figure, $v=\frac{k}{{z}}{r_a}r$.}
\label{BesselFunction} 
\end{figure*}

To obtain the estimation model for coupling noise, we first need to derive an approximate expression for the far-field WFE to establish the connection. Based on the derivation in Section \ref{se:2}, we know that the effects of aberrations on the far-field amplitude and phase can be expressed as a combination of Bessel functions. Fundamentally, the undistorted Gaussian beam contributes $J_1(v)$ to the real part. Therefore, the expression for the far-field WFE takes the form of $\sum\limits_{n}a_nJ_n(v)/\left[J_1(v)+\sum\limits_{m}b_mJ_m(v)\right]$. To ensure effective detection, the far-field WFE requires that the phase variation caused by a $10 \;\text{nrad}$ laser jitter must remain below $1\;\text{pm}/\sqrt{\text{Hz}}$ in the amplitude spectral density of the measurement noise\cite{chwalla2016design}. As shown in Fig. \ref{BesselFunction}, retaining terms up to $J_4(v)$ is sufficient to meet the required computational accuracy.
In addition to the transmitted WFE resulting from the telescope's design, factors such as mirror fabrication, structural assembly, and thermal effects due to the space environment during orbit also contribute to the transmitted WFE. Typically, the first 21 Zernike polynomials are adequate to describe these effects, as shown in Table \ref{ZernikeList}\cite{bely2003design}.
\begin{table*}
\abovetopsep=0pt
\aboverulesep=0pt
\belowrulesep=0pt
\belowbottomsep=0pt
	\begin{center}
			\caption{\textbf{The first 21 Zernike polynomials}}
		\begin{tabular}{c|c|c|c}
			\toprule[1.5pt]
			\textbf{Order} & \textbf{Aberration} & \textbf{Term} &  \textbf{Value} \\ 
			\midrule[1.5pt]
			1 & X-Tilt  & $Z_1^{1}$ & $\rho\cos{\phi}$\\
			\hline
			2 & Y-Tilt  & $Z_1^{-1}$ & $\rho\sin{\phi}$\\
			\hline
			3 & Defocus  & $Z_2^{0}$ & $2{\rho}^2-1$\\
			\hline	
			4 & $0^{\circ}$ Astigmatism  & $Z_2^{2}$ & ${\rho}^2\cos{2\phi}$\\
			\hline
			5 & $45^{\circ}$ Astigmatism  & $Z_2^{-2}$ & ${\rho}^2\sin{2\phi}$\\
			\hline
			6 & X-Coma  & $Z_3^{1}$ & $(3{\rho}^3-2\rho)\cos{\phi}$\\
			\hline
			7 & Y-Coma  & $Z_3^{-1}$ & $(3{\rho}^3-2\rho)\sin{\phi}$\\
			\hline
			8 & X-Trefoil  & $Z_3^{3}$ & ${\rho}^3\cos{3\phi}$\\
			\hline
			9 & Y-Trefoil  & $Z_3^{-3}$ & ${\rho}^3\sin{3\phi}$\\
			\hline
			10 & Spherical  & $Z_4^{0}$ & $6{\rho}^4-6{\rho}^2+1$\\
			\hline
			11 & X-$2^{nd}$ Astigmatism  & $Z_4^{2}$ & $(4{\rho}^4-3{\rho}^2)\cos{2\phi}$\\
			\hline	
			12 & Y-$2^{nd}$ Astigmatism  & $Z_4^{-2}$ & $(4{\rho}^4-3{\rho}^2)\sin{2\phi}$\\
			\hline
			13 & X-Tetrafoil  & $Z_4^{4}$ & ${\rho}^4\cos{4\phi}$\\
			\hline
			14 & Y-Tetrafoil  & $Z_4^{-4}$ & ${\rho}^4\sin{4\phi}$\\
			\hline
			15 & X-$2^{nd}$ Coma  & $Z_5^{1}$ & $(10{\rho}^5-12{\rho}^3+3{\rho})\cos{\phi}$\\
			\hline
			16 & Y-$2^{nd}$ Coma  & $Z_5^{-1}$ & $(10{\rho}^5-12{\rho}^3+3{\rho})\sin{\phi}$\\
			\hline
			17 & X-$2^{nd}$ Trefoil  & $Z_5^{3}$ & $(5{\rho}^5-4{\rho}^3)\cos{3\phi}$\\
			\hline
			18 & Y-$2^{nd}$ Trefoil  & $Z_5^{-3}$ & $(5{\rho}^5-4{\rho}^3)\sin{3\phi}$\\
			\hline
			19 & X-Pentafoil  & $Z_5^{5}$ & ${\rho}^5\cos{5\phi}$\\
			\hline
			20 & Y-Pentafoil  & $Z_5^{-5}$ & ${\rho}^5\sin{5\phi}$\\
			\hline
			21 & $2^{nd}$ Spherical  & $Z_6^{0}$ & $20{\rho}^6-30{\rho}^4+12{\rho}^2-1$\\
			\bottomrule[1.5pt]	
		\end{tabular}
			\label{ZernikeList}
	\end{center}
\end{table*}
\subsection{Transmitted wavefront error constraint} \label{sbse:4.1}
For clarity in this paper, we use Peak-to-Valley (P-V) error to quantify the transmitted WFE. Next, we provide an approximate constraint based on the physical image from Section \ref{se:3}. Previous studies have shown that the impact of $Z_2^{0}$ and $Z_2^{{\pm}2}$ on the far-field wavefront is significantly greater than that of higher-order aberrations\cite{kenny2020beam}. Therefore, it is reasonable to first impose an approximate constraint on the transmitted WFE using $Z_2^{0}$ or $Z_2^{{\pm}2}$. In this case, we choose $Z_2^{0}$ and derive an approximate expression for \eqref{FraunhoferAberration2} with only $Z_2^{0}$ aberration:
\begin{equation}\label{Z20U}
U(r, \psi, z)=U_0(r, \psi, z)+Z_2^{0}(r, \psi, z),
\end{equation}
\begin{equation}\label{U0}
\begin{aligned}
U_0(r, \psi, z)&=e^{-\frac{1}{2}}{\pi}^{\frac{1}{2}}(2\pi)\left\{I_{\frac{1}{2}}\frac{J_1(v)}{v}+3I_{\frac{3}{2}}\frac{J_3(v)}{v}\right\}\\
&=e^{-\frac{1}{2}}{\pi}^{\frac{1}{2}}(2\pi)\left\{{\sigma}_0\frac{J_1(v)}{v}+{\tau}_0\frac{J_3(v)}{v}\right\},
\end{aligned}
\end{equation}
where
\begin{gather*}\label{co00}
	\left({\sigma}_0,\;{\tau}_0\right)=\left(\num{0.587993},\;\num{0.28921}\right).
\end{gather*}
Here we let $I_{l+\frac{1}{2}}$ denote $I_{l+\frac{1}{2}}(\frac{1}{2})$. And
\begin{equation}\label{Z20}
\begin{aligned}
{Z_2^{0}(r, \psi, z)}_{1st+2nd+3rd}&=e^{-\frac{1}{2}}{\pi}^{\frac{1}{2}}(2\pi)\\
&\left\{ia_2^0\left[-I_{\frac{1}{2}}\frac{J_3(v)}{v}+(-3)I_{\frac{3}{2}}(\frac{1}{3})\frac{J_1(v)}{v}+5I_{\frac{5}{2}}(-\frac{2}{5})\frac{J_3(v)}{v}\right]+\right.\\
&\left.-\frac{(a_2^0)^2}{2}\left[I_{\frac{1}{2}}(\frac{1}{3})\frac{J_1(v)}{v}+(-3)I_{\frac{3}{2}}(-\frac{3}{5})\frac{J_3(v)}{v}+5I_{\frac{5}{2}}(\frac{2}{15})\frac{J_1(v)}{v}\right]+\right.\\
&\left.-i\frac{(a_2^0)^3}{6}\left[I_{\frac{1}{2}}(-\frac{3}{5})\frac{J_3(v)}{v}+(-3)I_{\frac{3}{2}}(\frac{1}{5})\frac{J_1(v)}{v}+5I_{\frac{5}{2}}(-\frac{12}{35})\frac{J_3(v)}{v}\right]\right\}\\
&=e^{-\frac{1}{2}}{\pi}^{\frac{1}{2}}(2\pi)\left\{\left[{\sigma}_2^0\frac{J_1(v)}{v}+{\tau}_2^0\frac{J_3(v)}{v}\right]+i\left[{\alpha}_2^0\frac{J_1(v)}{v}+{\beta}_2^0\frac{J_3(v)}{v}\right]\right\},
\end{aligned}
\end{equation}
where
\begin{gather*}\label{co20}
	\left({\sigma}_2^0,\;{\tau}_2^0\right)=(a_2^0)^2\left(-0.10119,\;-0.0867631\right),\\
	\left({\alpha}_2^0,\;{\beta}_2^0\right)=a_2^0\left(-0.0964035,-0.607138\right)+(a_2^0)^3\left(0.00964035,\;0.0615342\right).
\end{gather*}
Thus, the expression for the far-field WFE resulting from the single aberration $Z_2^{0}$ is:
\begin{equation}\label{WFE_Z20}
{\delta}{\Theta}_2^0(r, \psi, z)=\frac{{\alpha}_2^0J_1(v)+{\beta}_2^0J_3(v)}{({\sigma}_0+{\sigma}_2^0)J_1(v)+({\tau}_0+{\tau}_2^0)J_3(v)}-\frac{{\alpha}_2^0J_1(0)+{\beta}_2^0J_3(0)}{({\sigma}_0+{\sigma}_2^0)J_1(0)+({\tau}_0+{\tau}_2^0)J_3(0)}.
\end{equation}

\begin{figure*}[htbp]
	\begin{center}
		\includegraphics[width=0.8\textwidth]{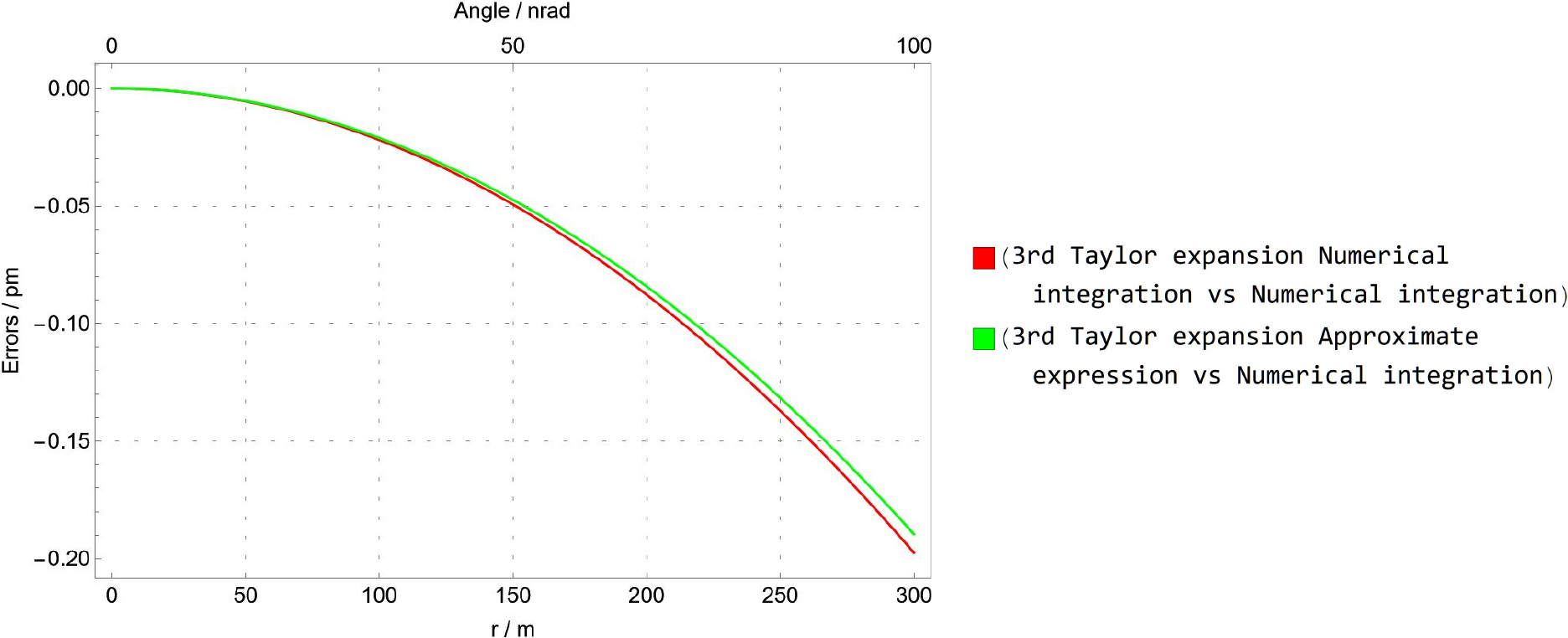}
	\end{center}
	\caption{The error of the numerical integration obtained from the third-order Taylor expansion of $e^{i\varOmega_a(\rho,\theta)}$ compared to the direct numerical integration of \eqref{FraunhoferAberration2}, along with the error of the approximate expression relative to the direct numerical integration of \eqref{FraunhoferAberration2}.}
\label{ErrorLevel} 
\end{figure*}

By comparing the result of the approximate expression \eqref{WFE_Z20} with the numerical integration result from \eqref{FraunhoferAberration2}, we find that, at a transmitted WFE of $\lambda/4$ (P-V), the error within a $100\; \text{nrad}$ angle range is approximately 0.18 pm. Additionally, we compare the numerical integration results obtained from the third-order Taylor expansion of $e^{i\varOmega_a(\rho,\theta)}$ in \eqref{FraunhoferAberration2} with those obtained from directly integrating \eqref{FraunhoferAberration2}. The two error levels are closely aligned, as shown in Fig. \ref{ErrorLevel}. Therefore, the approximate expression is validated.

\begin{figure*}[htbp]
	\begin{center}
		\includegraphics[width=0.7\textwidth]{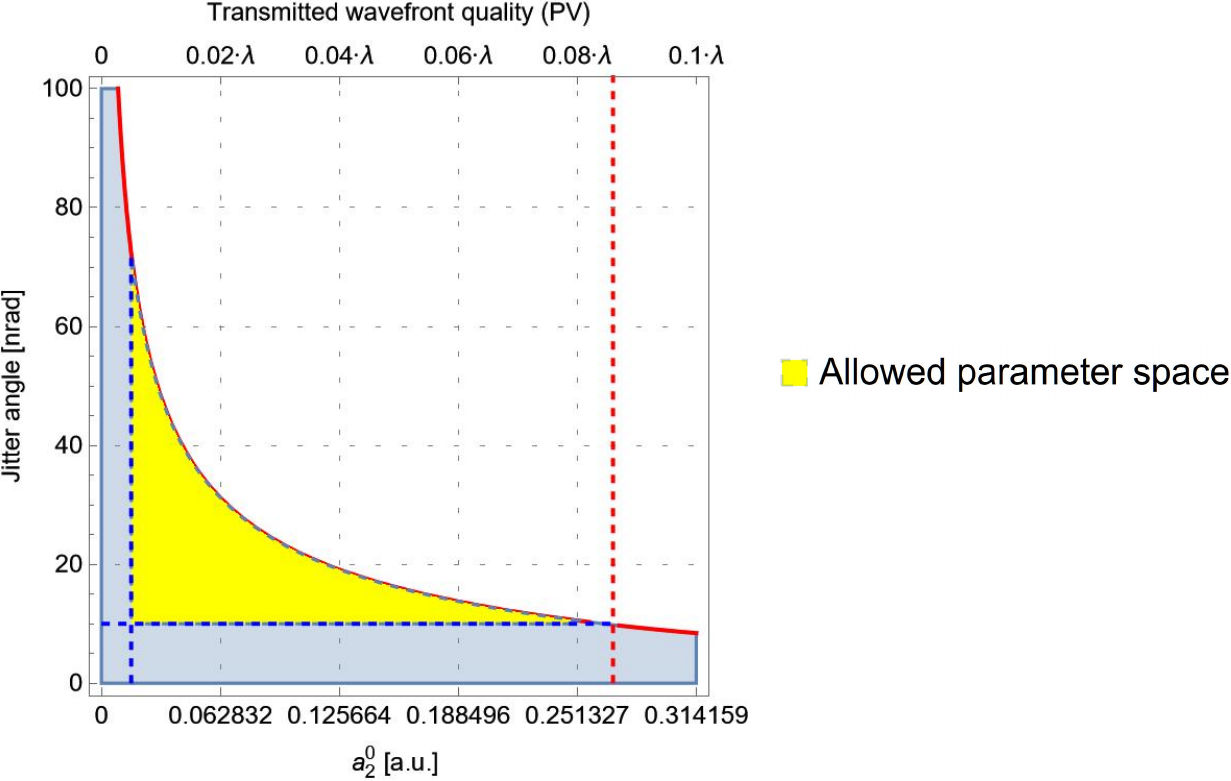}
	\end{center}
	\caption{The constraints on laser jitter and transmitted WFE for a single aberration, specifically defocus, under a far-field WFE of $1 \text{pm}$ and a static pointing angle of $10 \text{nrad}$.}
\label{ParaWindow} 
\end{figure*}

A parameter space can be defined for transmitted WFE, static pointing angle, and laser jitter, based on the requirement that the far-field WFE remains within $1\text{pm}$. Figure \ref{ParaWindow} illustrates the constraints on laser jitter and transmitted WFE for a single aberration $Z_2^{0}$, with a static pointing angle of $10\text{nrad}$. Given a laser jitter of $10\text{nrad}$, the transmitted WFE for defocus must be less than approximately $\lambda/12$. To generalize the model more general, we propose a broader requirement that the transmitted WFE should be less than approximately $\lambda/10$. We require that, under this transmitted WFE, the error of the approximate expression compared to the numerical integration is approximately at the level of $0.1\;\text{pm}$ within a region of $100\;\text{nrad}$.

\subsection{The real and imaginary parts of $U(r, \psi, z)$} \label{sbse:4.2}
Next, we derive the real and imaginary parts of $U(r, \psi, z)$. First, we summarize the aberration terms and their higher-order coupling terms that correspond to either the real or imaginary part of $U(r, \psi, z)$. In the third-order Taylor expansion of $e^{i\varOmega_a(\rho,\theta)}$:
\begin{equation}
	e^{i\varOmega_a(\rho,\theta)}=1+i\varOmega_a-\frac{1}{2}{\varOmega_a}^2-i\frac{1}{6}{\varOmega_a}^3,
	\label{Expansion}
\end{equation}
 Odd-order terms introduce $i$ while the even-order terms do not. For first-order terms, all Zernike terms with odd $m$ introduce an additional $i$ through \eqref{BesselIntegrals}, classifying them as belonging to the real part of $U(r, \psi, z)$. In contrast, Zernike terms with even $m$ are categorized as belonging to the imaginary part. For the second-order terms, the coupling terms between the Zernike terms with odd $m$ do not introduce $i$ in the integral by applying Product-to-Sum Identities and \eqref{BesselIntegrals}. Thus, they also belong to the real part of $U(r, \psi, z)$. Similarly, after analyzing the coupling of third-order terms, we summarize our conclusions in Table \ref{ReAndIm}.  
\begin{table*}
\abovetopsep=0pt
\aboverulesep=0pt
\belowrulesep=0pt
\belowbottomsep=0pt
	\begin{center}
		\begin{tabular}{c|c|c}
			\toprule[1.5pt]
			  & Real part of $U(r, \psi, z)$ &  Imaginary part of $U(r, \psi, z)$\\ 
		\midrule[1.5pt]
  First order & ${Z^{2\alpha+1}_{\gamma}}$ & ${Z^{2\beta}_{{\gamma}{'}}}$\\
  Second order & ${Z^{2{\alpha}_1+1}_{{\gamma}_1}}{Z^{2{\alpha}_2+1}_{{\gamma}_2}}\quad{Z^{2{\beta}_1}_{{{\gamma}_1}{'}}}{Z^{2{\beta}_2}_{{{\gamma}_2}{'}}}$ & ${Z^{2\alpha+1}_{\gamma}}{Z^{2\beta}_{{\gamma}{'}}}$\\
    Third order & ${Z^{2{\alpha}_1+1}_{{\gamma}_1}}{Z^{2{\alpha}_2+1}_{{\gamma}_2}}{Z^{2{\alpha}_3+1}_{{\gamma}_3}}\quad{Z^{2{\alpha}+1}_{{\gamma}}}{Z^{2{\beta}_1}_{{{\gamma}_1}{'}}}{Z^{2{\beta}_2}_{{{\gamma}_2}{'}}}$ & ${Z^{2{\alpha}_1+1}_{{\gamma}_1}}{Z^{2{\alpha}_2+1}_{{\gamma}_2}}{Z^{2{\beta}}_{{{\gamma}}{'}}}\quad{Z^{2{\beta}_1}_{{{\gamma}_1}{'}}}{Z^{2{\beta}_2}_{{{\gamma}_2}{'}}}{Z^{2{\beta}_3}_{{{\gamma}_3}{'}}}$\\
			\bottomrule[1.5pt]
		\end{tabular}
		\caption{The aberration terms and their higher-order coupling terms corresponding to the real and imaginary parts of $U(r, \psi, z)$, where ${Z^{2\alpha+1}_{\gamma}}$ denotes the Zernike terms $Z^m_n$ with odd $m$, and ${Z^{2\beta}_{{\gamma}{'}}}$ denotes those with even $m$.}
\label{ReAndIm}
	\end{center}
\end{table*}
These terms contribute differently to the far-field WFE ${\delta}{\Theta}(r, \psi, z)$, and we can discard those that have negligible contributions. Based on the subsequent comparison results between A.E. and N.I., retaining only the first two orders is sufficient to meet the error requirements. For the third-order term, we retain only the contribution from $Z_2^0$.

\begin{itemize}
\item[$\blacksquare$]${Z^{2\beta}_{{\gamma}{'}}}$

For Zernike terms where $m$ is even, Similar to the derivation of $Z_2^{0}$ in Subsection \ref{sbse:4.1}, we obtain the expressions for each even single aberration term:
\begin{equation}\label{Zn2}
\begin{aligned}
	\begin{pmatrix}
			Z_n^{2} \\
			Z_n^{-2} \\
		\end{pmatrix}
(r, \psi, z)=e^{-\frac{1}{2}}{\pi}^{\frac{1}{2}}(2\pi)\left\{-i
		\begin{pmatrix}
			\cos2\psi \\
			\sin2\psi \\
		\end{pmatrix}
		\left[{\alpha}_n^{\pm2}\frac{J_1(v)}{v}+{\beta}_n^{\pm2}\frac{J_3(v)}{v}\right]\right\},
\end{aligned}
\end{equation}
\begin{equation}\label{Zn0}
\begin{aligned}
Z_n^{0}(r, \psi, z)=e^{-\frac{1}{2}}{\pi}^{\frac{1}{2}}(2\pi)\left\{i\left[{\alpha}_n^{0}\frac{J_1(v)}{v}+{\beta}_n^{0}\frac{J_3(v)}{v}\right]\right\},
\end{aligned}
\end{equation}
where
\begin{gather*}\label{coEven}
\left({\alpha}_2^{\pm2},\;{\beta}_2^{\pm2}\right)={a_2^{\pm2}}\left(0,\;\num{-0.448174}\right),\\
\left({\alpha}_4^{0},\;{\beta}_4^{0}\right)={a_4^0}\left(\num{0.00957224},\; \num{0.115684}\right),\\
\left({\alpha}_4^{\pm2},\;{\beta}_4^{\pm2}\right)={a_4^{\pm2}}\left(0,\;\num{0.0724048}\right),\\
\left({\alpha}_6^{0},\;{\beta}_6^{0}\right)=a_6^0\left(0,\;\num{-0.0123072}\right).
\end{gather*}
Since the lowest-order term in the Bessel function expansion of $Z_4^{\pm4}$ is $J_5(v)$, we can directly discard $Z_4^{\pm4}$. We compared the WFE differences between the Approximate Expression (A.E.) and the numerical integration (N.I.) of \eqref{FraunhoferAberration2}, as shown in Fig. \ref{1stEven}. In this calculation, $Z_2^0$ only considers the first-order expansion, specifically $({\sigma}_2^0,\;{\tau}_2^0)=(0,\;0)$ and $({\alpha}_2^0,\;{\beta}_2^0)=(-0.0964035a_2^0,\;-0.607138a_2^0)$.
\begin{figure}
     \centering
     \begin{subfigure}[b]{0.3\textwidth}
         \centering
         \caption{$Z_2^0$}
         \includegraphics[width=1\textwidth]{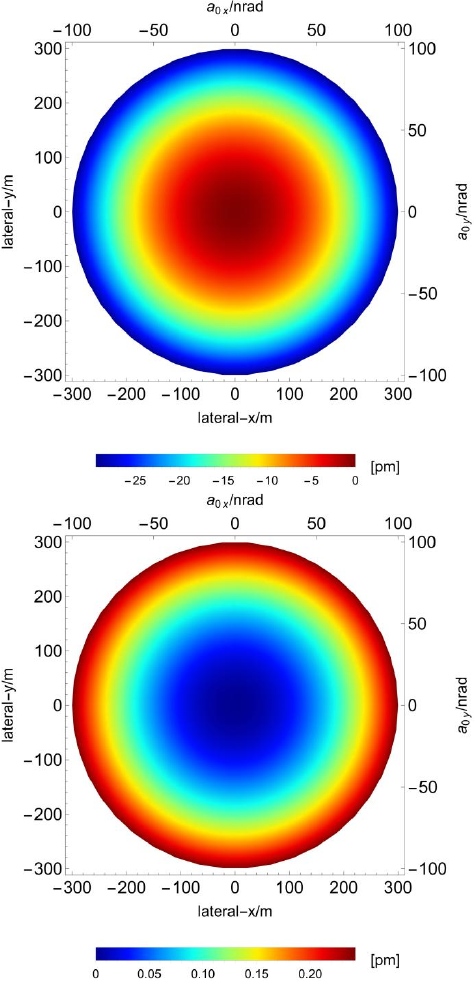}
         \label{fig:Z20}
     \end{subfigure}
     \begin{subfigure}[b]{0.3\textwidth}
         \centering
         \caption{$Z_4^0$}
         \includegraphics[width=1\textwidth]{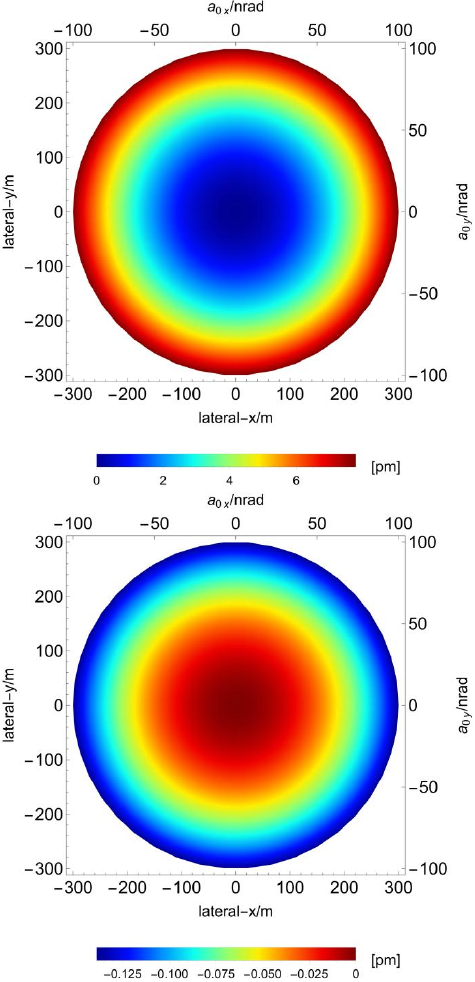}
         \label{fig:Z40}
     \end{subfigure}    
          \begin{subfigure}[b]{0.3\textwidth}
         \centering
         \caption{$Z_6^0$}
         \includegraphics[width=1\textwidth]{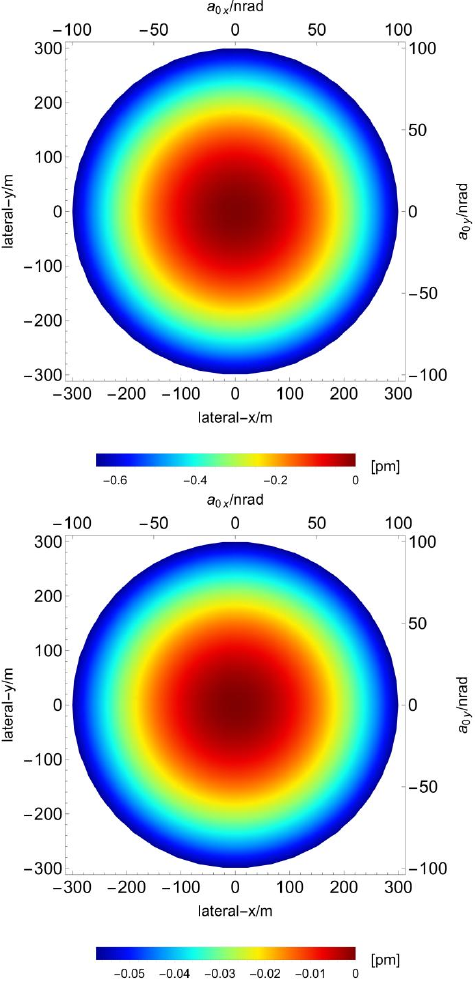}
         \label{fig:Z60}        
     \end{subfigure}
     
          \begin{subfigure}[b]{0.3\textwidth}
         \centering
         \caption{$Z_2^{\pm2}$}
         \includegraphics[width=1\textwidth]{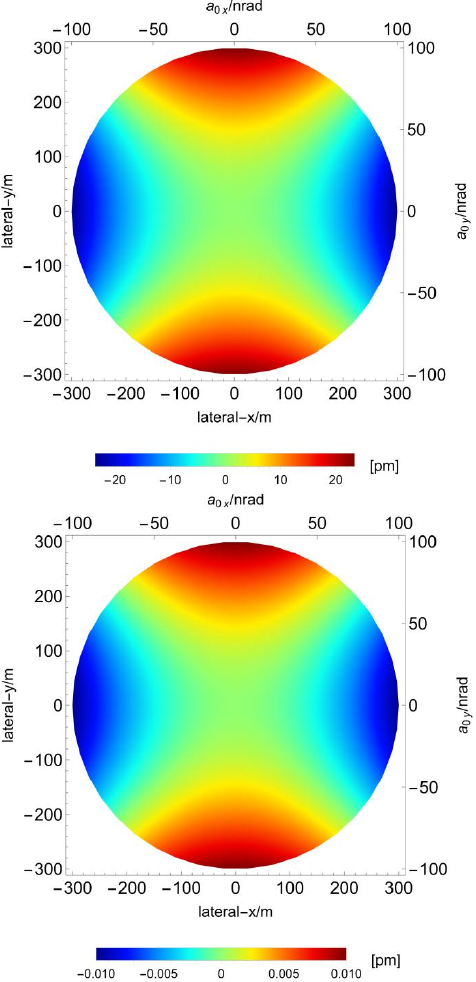}
         \label{fig:Z22}        
     \end{subfigure}
          \begin{subfigure}[b]{0.3\textwidth}
         \centering
         \caption{$Z_4^{\pm2}$}
         \includegraphics[width=1\textwidth]{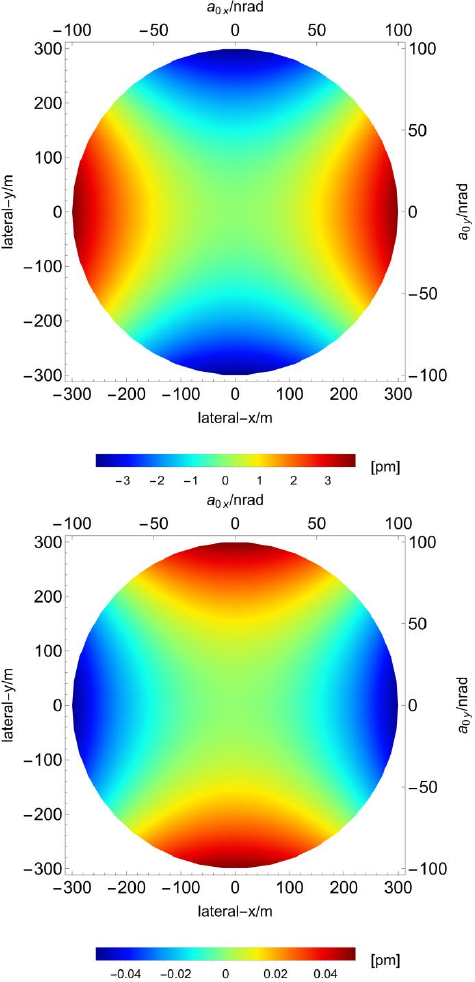}
         \label{fig:Z42}
     \end{subfigure}
        \caption{The far-field WFE of ${Z^{2\beta}_{{\gamma}{'}}}$ (the upper half of each subfigure) and the WFE difference between the Approximate Expression (A.E.) and the numerical integration (N.I.) of \eqref{FraunhoferAberration2} (the lower half of each subfigure). The transmitted WFEs for each aberration are constrained to $\lambda/10$, with  $a_4^0=0.418879$ and the other coefficients set to 0.314159. The terms of $Z_2^{\pm2}$ and $Z_4^{\pm2}$ only display $Z_2^{2}$ and $Z_4^{2}$.}
        \label{1stEven}
\end{figure}
Typically, lower-order Zernike aberrations contribute more to the transmitted WFE than higher-order Zernike aberrations. By considering the contributions of each aberration to the transmitted WFE and the far-field WFE, as well as the errors in the A.E., we retain only the contribution from $Z_2^0$.\\
\end{itemize}

\begin{itemize}
\item[$\blacksquare$]${Z^{2\alpha+1}_{\gamma}}$

For ${Z^{2\alpha+1}_{\gamma}}$, similarly, we obtain:
\begin{equation}\label{Zn1}
\begin{aligned}
	\begin{pmatrix}
			Z_n^{1} \\
			Z_n^{-1} \\
		\end{pmatrix}
(r, \psi, z)=e^{-\frac{1}{2}}{\pi}^{\frac{1}{2}}(2\pi)\left\{
		\begin{pmatrix}
			\cos\psi \\
			\sin\psi \\
		\end{pmatrix}
		\left[{\sigma}_n^{\pm1}\frac{J_2(v)}{v}+{\tau}_n^{\pm1}\frac{J_4(v)}{v}\right]\right\},
\end{aligned}
\end{equation}
\begin{equation}\label{Z33}
\begin{aligned}
	\begin{pmatrix}
			Z_n^{3} \\
			Z_n^{-3} \\
		\end{pmatrix}
(r, \psi, z)=e^{-\frac{1}{2}}{\pi}^{\frac{1}{2}}(2\pi)\left\{-
	\begin{pmatrix}
			\cos3\psi \\
			\sin3\psi \\
		\end{pmatrix}
\left[{\sigma}_n^{\pm3}\frac{J_2(v)}{v}+{\tau}_n^{\pm3}\frac{J_4(v)}{v}\right]\right\},
\end{aligned}
\end{equation}
where
\begin{gather*}\label{coEven}
\left({\sigma}_1^{\pm1},\;{\tau}_1^{\pm1}\right)=a_1^{\pm1}\left(\num{0.49159},\;\num{0.173662}\right),\\
\left({\sigma}_3^{\pm1},\;{\tau}_1^{\pm1}\right)=a_3^{\pm1}\left(\num{-0.0868312},\;\num{-0.578285}\right),\\
\left({\sigma}_3^{\pm3},\;{\tau}_3^{\pm3}\right)=a_3^{\pm3}\left(0,\;\num{0.424039}\right),\\
\left({\sigma}_5^{\pm1},\;{\tau}_5^{\pm1}\right)=a_5^{\pm1}\left(\num{0.00957224},\;\num{0.112949}\right),\\
\left({\sigma}_5^{\pm3},\;{\tau}_5^{\pm3}\right)=a_5^{\pm3}\left(0,\;\num{0.0200039}\right).
\end{gather*}
We can discard $Z_5^{\pm5}$ as well. Since ${Z^{2\alpha+1}_{\gamma}}$ does not affect the imaginary part in the absence of other coupled aberrations, we cannot directly measure these terms or their errors using far-field WFE, as we did with ${Z^{2\beta}_{{\gamma}{'}}}$. In this context, we use $Z_2^0$ to estimate their contributions by adding a term ${Z^{2\alpha+1}_{\gamma}}$ solely to the real part during the calculation and comparing the difference between this “WFE” and the WFE of $Z_2^0$, as shown in Fig. \ref{1stOdd}. It is important to note that this does not represent the true level of the far-field WFE for the combined aberration.
\begin{figure}
     \centering
     \begin{subfigure}[b]{0.3\textwidth}
         \centering
         \caption{$Z_1^{\pm1}$}
         \includegraphics[width=1\textwidth]{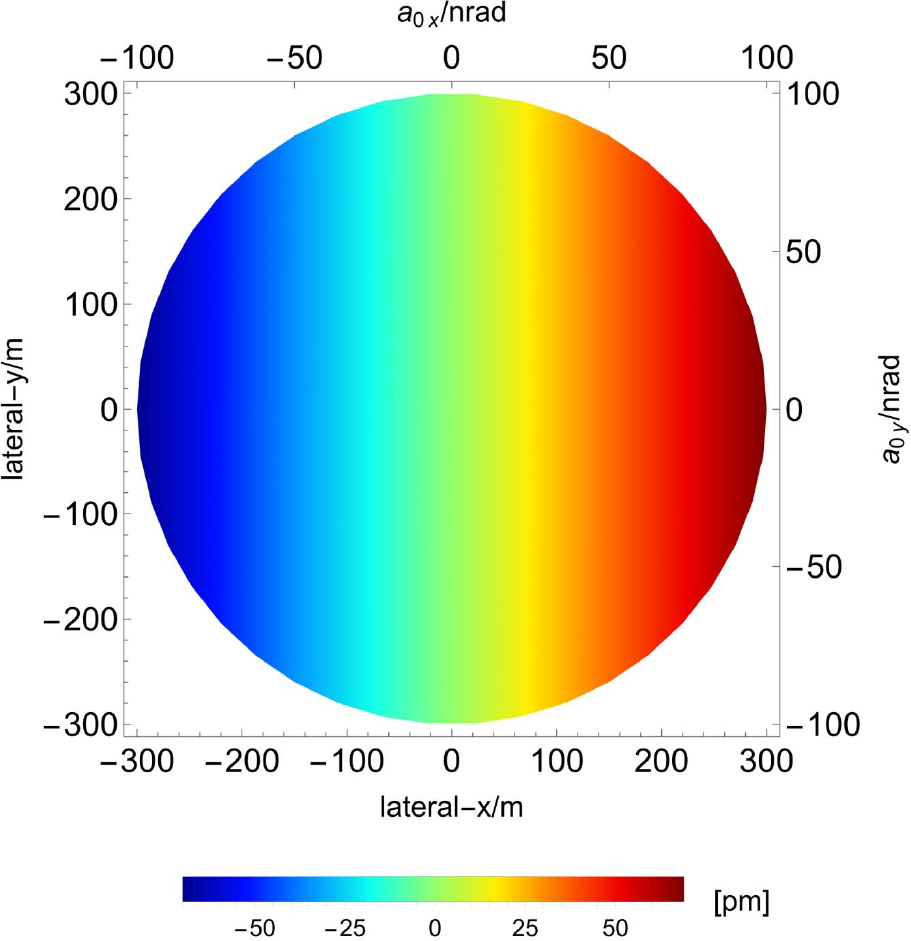}
         \label{fig:Z11}
     \end{subfigure}
     \begin{subfigure}[b]{0.3\textwidth}
         \centering
         \caption{$Z_3^{\pm1}$}
         \includegraphics[width=1\textwidth]{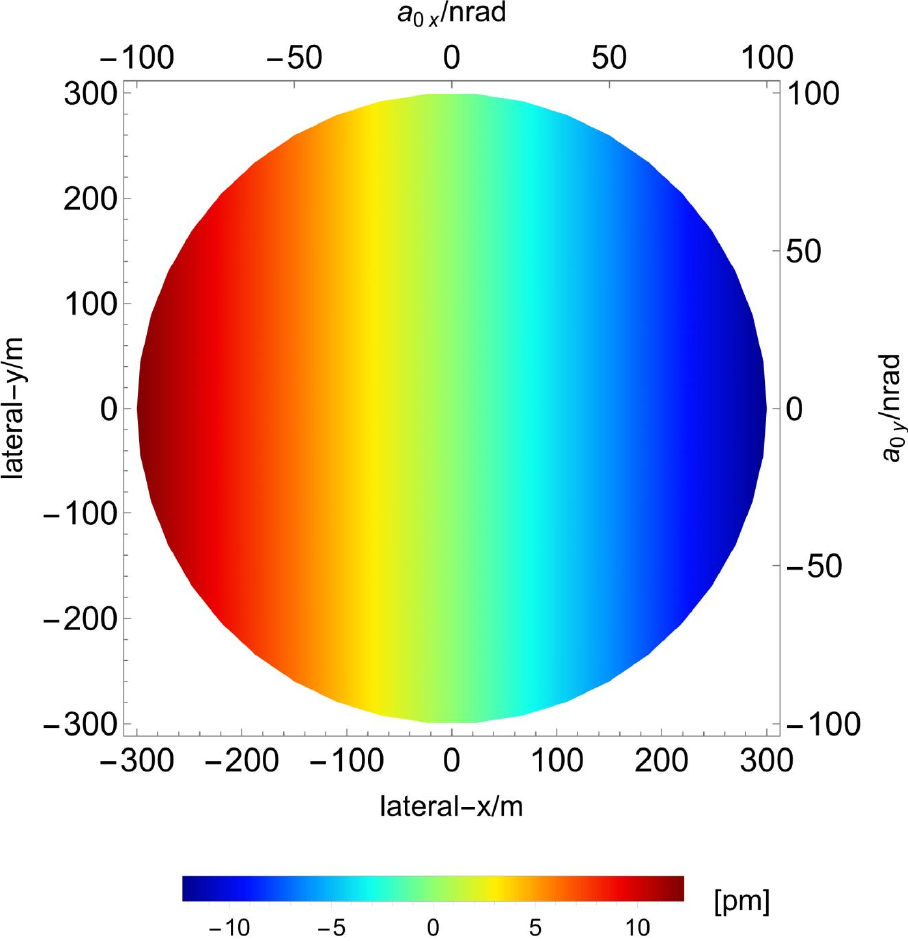}
         \label{fig:Z31}
     \end{subfigure}
          \begin{subfigure}[b]{0.3\textwidth}
         \centering
         \caption{$Z_5^{\pm1}$}
         \includegraphics[width=1\textwidth]{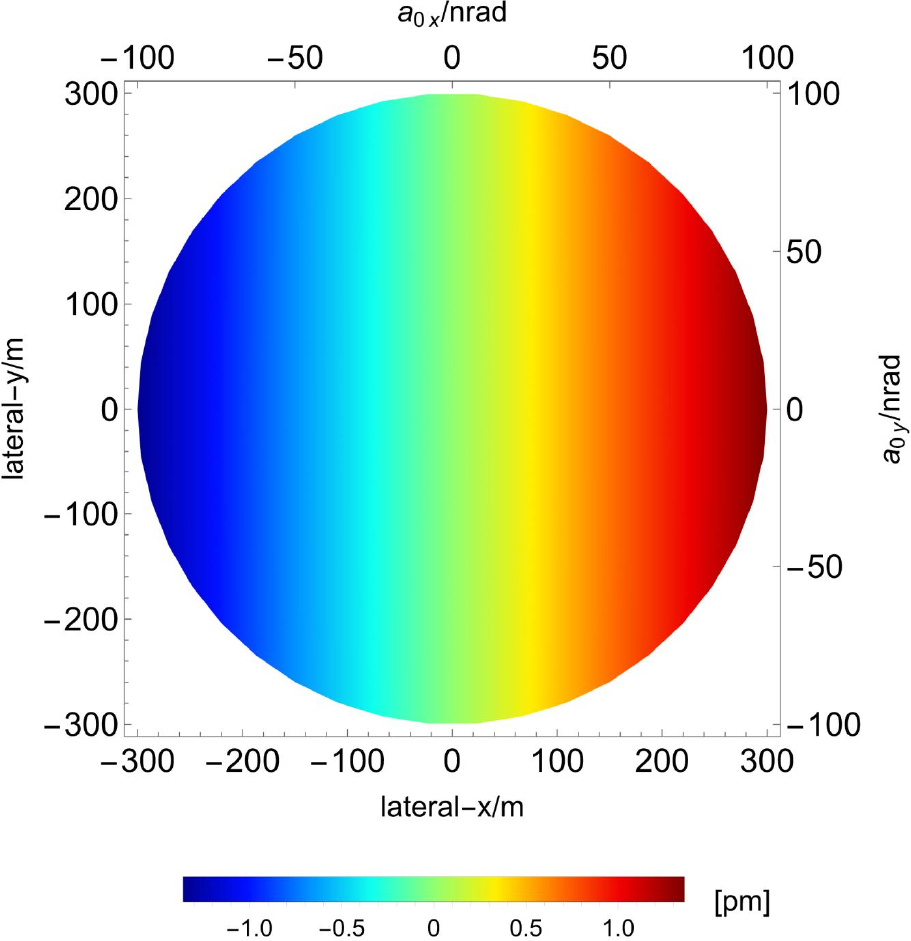}
         \label{fig:Z51}        
     \end{subfigure}
          \begin{subfigure}[b]{0.3\textwidth}
         \centering
         \caption{$Z_3^{\pm3}$}
         \includegraphics[width=1\textwidth]{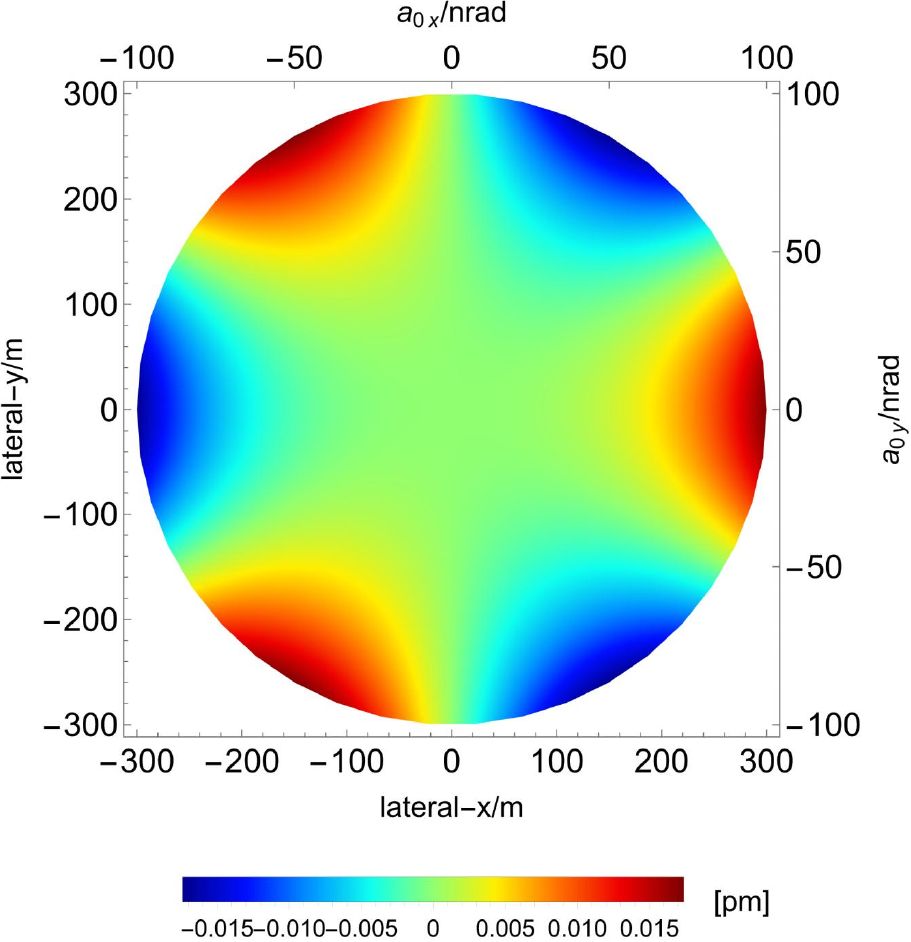}
         \label{fig:Z33}        
     \end{subfigure}
          \begin{subfigure}[b]{0.3\textwidth}
         \centering
         \caption{$Z_5^{\pm3}$}
         \includegraphics[width=1\textwidth]{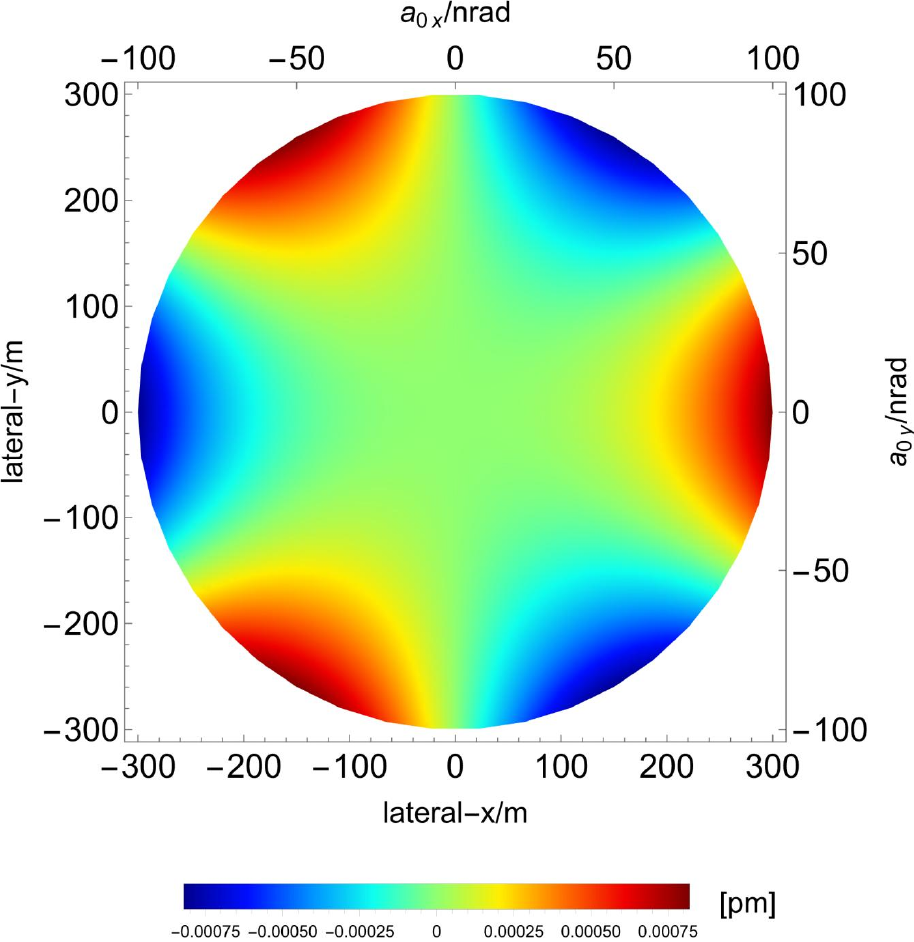}
         \label{fig:Z53}
     \end{subfigure}
        \caption{The Difference between $"{WFE}_{{Z^{2\alpha+1}_{\gamma}}+Z_2^0}"$ and ${WFE}_{Z_2^0}$, calculated using A.E.. $a^0_2$ and $a^{2\alpha+1}_{\gamma}$ are set to be 0.314159. The terms of $Z_{\gamma}^{\pm1}$ and $Z_{\gamma}^{\pm3}$ only display $Z_{\gamma}^{1}$ and $Z_{\gamma}^{3}$.}
        \label{1stOdd}
\end{figure}

We can thus conclude that $Z_3^{\pm3}$ and $Z_5^{\pm3}$ along with their higher-order terms, can be discarded, while $Z_1^{\pm1}$, $Z_3^{\pm1}$, $Z_5^{\pm1}$, and their higher-order couplings need to be retained.\\
\end{itemize}

\begin{itemize}
\item[$\blacksquare$]${Z^{2\alpha+1}_{\gamma}}{Z^{2\beta}_{{\gamma}{'}}}$

For ${Z^{2\alpha+1}_{\gamma}}{Z^{2\beta}_{{\gamma}{'}}}$, we obtain:
\begin{equation}\label{Zn1Zn'0}
	\begin{pmatrix}
			Z_n^{1}Z_{n{'}}^{0} \\
			Z_n^{-1}Z_{n{'}}^{0} \\
		\end{pmatrix}
		(r, \psi, z)=e^{-\frac{1}{2}}{\pi}^{\frac{1}{2}}(2\pi)\left\{i
	\begin{pmatrix}
			\cos\psi \\
			\sin\psi \\
		\end{pmatrix}
	\left[{\alpha}_{n;n{'}}^{\pm1;0}\frac{J_2(v)}{v}+{\beta}_{n;n{'}}^{\pm1;0}\frac{J_4(v)}{v}\right]\right\},
\end{equation}
\begin{equation}\label{Zn1Zn'pm2}
\begin{aligned}
	&\begin{pmatrix}
			Z_n^{1}Z_{n{'}}^{2} &  Z_n^{1}Z_{n{'}}^{-2}\\
			Z_n^{-1}Z_{n{'}}^{2} & Z_n^{-1}Z_{n{'}}^{-2}\\
		\end{pmatrix}
		(r, \psi, z)=e^{-\frac{1}{2}}{\pi}^{\frac{1}{2}}(2\pi)\\
		&\left\{\frac{i}{2}
	\begin{pmatrix}
			\cos\psi &  \sin\psi\\
			-\sin\psi & \cos\psi\\
		\end{pmatrix}
			\left[{{\alpha}_1}_{n;n{'}}^{\pm1;\pm2}\frac{J_2(v)}{v}+{{\beta}_1}_{n;n{'}}^{\pm1;\pm2}\frac{J_4(v)}{v}\right]-\frac{i}{2}
		\begin{pmatrix}
			\cos3\psi &  \sin3\psi\\
			\sin3\psi & -\cos3\psi\\
		\end{pmatrix}				
			\left[{{\alpha}_2}_{n;n{'}}^{\pm1;\pm2}\frac{J_2(v)}{v}+{{\beta}_2}_{n;n{'}}^{\pm1;\pm2}\frac{J_4(v)}{v}\right]\right\},	
\end{aligned}
\end{equation}
\begin{equation}\label{Zn1Zn'pm4}
\begin{aligned}
	&\begin{pmatrix}
			Z_n^{1}Z_{n{'}}^{4} &  Z_n^{1}Z_{n{'}}^{-4}\\
			Z_n^{-1}Z_{n{'}}^{4} & Z_n^{-1}Z_{n{'}}^{-4}\\
		\end{pmatrix}
		(r, \psi, z)=e^{-\frac{1}{2}}{\pi}^{\frac{1}{2}}(2\pi)
		\left\{-\frac{i}{2}
		\begin{pmatrix}
			\cos3\psi &  \sin3\psi\\
			-\sin3\psi & \cos3\psi\\
		\end{pmatrix}				
			\left[{{\alpha}}_{n;n{'}}^{\pm1;\pm4}\frac{J_2(v)}{v}+{{\beta}}_{n;n{'}}^{\pm1;\pm4}\frac{J_4(v)}{v}\right]\right\},	
\end{aligned}
\end{equation}
\begin{equation}\label{Zn3Zn'0}
	\begin{pmatrix}
			Z_n^{3}Z_{n{'}}^{0} \\
			Z_n^{-3}Z_{n{'}}^{0} \\
		\end{pmatrix}
		(r, \psi, z)=e^{-\frac{1}{2}}{\pi}^{\frac{1}{2}}(2\pi)\left\{-i
	\begin{pmatrix}
			\cos3\psi \\
			\sin3\psi \\
		\end{pmatrix}
	\left[{\alpha}_{n;n{'}}^{\pm3;0}\frac{J_2(v)}{v}+{\beta}_{n;n{'}}^{\pm3;0}\frac{J_4(v)}{v}\right]\right\},	
\end{equation}
\begin{equation}\label{Zn3Zn'pm2}
\begin{aligned}
	&\begin{pmatrix}
			Z_n^{3}Z_{n{'}}^{2} &  Z_n^{3}Z_{n{'}}^{-2}\\
			Z_n^{-3}Z_{n{'}}^{2} & Z_n^{-3}Z_{n{'}}^{-2}\\
		\end{pmatrix}
		(r, \psi, z)=e^{-\frac{1}{2}}{\pi}^{\frac{1}{2}}(2\pi)
		\left\{\frac{i}{2}
	\begin{pmatrix}
			\cos\psi &  -\sin\psi\\
			\sin\psi & \cos\psi\\
		\end{pmatrix}
			\left[{{\alpha}}_{n;n{'}}^{\pm3;\pm2}\frac{J_2(v)}{v}+{{\beta}}_{n;n{'}}^{\pm3;\pm2}\frac{J_4(v)}{v}\right]\right\},	
\end{aligned}
\end{equation}
\begin{equation}\label{Zn3Zn'pm4}
\begin{aligned}
	&\begin{pmatrix}
			Z_n^{3}Z_{n{'}}^{4} &  Z_n^{3}Z_{n{'}}^{-4}\\
			Z_n^{-3}Z_{n{'}}^{4} & Z_n^{-3}Z_{n{'}}^{-4}\\
		\end{pmatrix}
		(r, \psi, z)=e^{-\frac{1}{2}}{\pi}^{\frac{1}{2}}(2\pi)
		\left\{\frac{i}{2}
		\begin{pmatrix}
			\cos\psi &  \sin\psi\\
			-\sin\psi & \cos\psi\\
		\end{pmatrix}				
			\left[{{\alpha}}_{n;n{'}}^{\pm3;\pm4}\frac{J_2(v)}{v}+{{\beta}}_{n;n{'}}^{\pm3;\pm4}\frac{J_4(v)}{v}\right]\right\},	
\end{aligned}
\end{equation}
\begin{equation}\label{Zn5Zn'pm2}
\begin{aligned}
	&\begin{pmatrix}
			Z_n^{5}Z_{n{'}}^{2} &  Z_n^{5}Z_{n{'}}^{-2}\\
			Z_n^{-5}Z_{n{'}}^{2} & Z_n^{-5}Z_{n{'}}^{-2}\\
		\end{pmatrix}
		(r, \psi, z)=e^{-\frac{1}{2}}{\pi}^{\frac{1}{2}}(2\pi)
		\left\{-\frac{i}{2}
	\begin{pmatrix}
			\cos3\psi &  -\sin3\psi\\
			\sin3\psi & \cos3\psi\\
		\end{pmatrix}
			\left[{{\alpha}}_{n;n{'}}^{\pm5;\pm2}\frac{J_2(v)}{v}+{{\beta}}_{n;n{'}}^{\pm5;\pm2}\frac{J_4(v)}{v}\right]\right\},	
\end{aligned}
\end{equation}
\begin{equation}\label{Zn5Zn'pm4}
\begin{aligned}
	&\begin{pmatrix}
			Z_n^{5}Z_{n{'}}^{4} &  Z_n^{5}Z_{n{'}}^{-4}\\
			Z_n^{-5}Z_{n{'}}^{4} & Z_n^{-5}Z_{n{'}}^{-4}\\
		\end{pmatrix}
		(r, \psi, z)=e^{-\frac{1}{2}}{\pi}^{\frac{1}{2}}(2\pi)
		\left\{\frac{i}{2}
	\begin{pmatrix}
			\cos\psi &  -\sin\psi\\
			\sin\psi & \cos\psi\\
		\end{pmatrix}
			\left[{{\alpha}}_{n;n{'}}^{\pm5;\pm4}\frac{J_2(v)}{v}+{{\beta}}_{n;n{'}}^{\pm5;\pm4}\frac{J_4(v)}{v}\right]\right\}.	
\end{aligned}
\end{equation}
And their coefficients are listed in Table \ref{2ndcoefficients}.
\begin{table*}
\abovetopsep=0pt
\aboverulesep=0pt
\belowrulesep=0pt
\belowbottomsep=0pt
\renewcommand\arraystretch{2}
	\begin{center}
			\setlength{\tabcolsep}{1mm}
		\begin{tabular}[width=0.7\textwidth]{c|c c c}
			\toprule[1.5pt]
			 & $Z_2^{0}$ &  $Z_2^{\pm2}$ &  $Z_4^{0}$  \\ 
			\midrule[1.5pt]
			$Z_1^{\pm1}$ & $(\num{-0.105976},\;\num{-0.327636})$ & \makecell[c]{${(\num{-0.298783},\;\num{-0.0769866})}_1$\\${(-0.,\;\num{0.424039})}_2$} & $(\num{-0.0289891},\;\num{-0.163544})$\\
			\hline
			$Z_3^{\pm1}$  &  $(\num{-0.163818},\;\num{0.0871047})$ & \makecell[c]{${(\num{-0.0384933},\;\num{-0.24559})}_1$\\${(-0.,\;\num{0.133188})}_2$} & $(\num{-0.0817722},\;\num{-0.0775725})$\\
			\hline
			$Z_3^{\pm3}$ &  $(-0.,\;\num{0.230138})$ & $(\num{-0.21202},\;\num{-0.133188})$ & $(-0.,\;\num{0.0555374})$\\
			\hline
			$Z_5^{\pm1}$  & $(\num{-0.034459},\;\num{-0.235119})$ & \makecell[c]{${(\num{-0.0124434},\;\num{-0.0610851})}_1$\\${(-0.,\;\num{0.00377052})}_2$} & $(\num{-0.0955441},\;\num{0.0270358})$\\
			\hline
			$Z_5^{\pm3}$ &  $(-0.,\;\num{0.100721})$ & $(\num{-0.0303566},\;\num{-0.181438})$ & $(-0.,\;\num{0.146471})$\\
			\hline
			$Z_5^{\pm5}$ &  $\backslash$ & $(-0.,\;\num{0.265672})$ & $\backslash$\\
			\midrule[1.5pt]
			\midrule[1.5pt]
			&  $Z_4^{\pm2}$ &  $Z_4^{\pm4}$ &  $Z_0^{6}$\\
			\midrule[1.5pt]
			$Z_1^{\pm1}$ &  \makecell[c]{${(\num{-0.0482698},\;\num{-0.301791})}_1$\\${(-0.,\;\num{0.0362372})}_2$} & $(-0.,\;\num{0.327089})$ & $(\num{-0.00410239},\;\num{0.0413741})$\\
			\hline
			$Z_3^{\pm1}$  &  \makecell[c]{${(\num{-0.150896},\;\num{0.0882182})}_1$\\${(-0.,\;\num{0.171788})}_2$} & $(-0.,\;\num{0.142838})$ & $(\num{-0.0206871},\;\num{-0.130511})$\\
			\hline
			$Z_3^{\pm3}$  &  $(\num{-0.0181186},\;\num{-0.171788})$ & $(\num{-0.163544},\;\num{-0.142838})$ & $(-0.,\;\num{-0.000240703})$\\
			\hline
			$Z_5^{\pm1}$  &  \makecell[c]{${(\num{-0.0448712},\;\num{-0.118361})}_1$\\${(-0.,\;\num{0.0992818})}_2$} & $(-0.,\;\num{0.0276483})$ & $(\num{-0.0610328},\;\num{-0.0350193})$\\
			\hline
			$Z_5^{\pm3}$  &  $(\num{-0.131078},\;\num{0.0499303})$ & $(\num{-0.0100019},\;\num{-0.123596})$ & $(-0.,\;\num{0.058526})$\\
			\hline
			$Z_5^{\pm5}$ &  $(-0.,\;\num{0.096762})$ & $(\num{-0.132836},\;\num{-0.138989})$ & $\backslash$\\
			\bottomrule[1.5pt]
		\end{tabular}
		\caption{The coefficient list of $Z^{2\alpha+1}_{\gamma}Z^{2\beta}_{{\gamma}{'}}$. Each term should be multiplied by the corresponding Zernike coefficients $a^{2\alpha+1}_{\gamma}$ and $a^{2\beta}_{{\gamma}{'}}$. "${(\quad)}_1$" and "${(\quad)}_2$" correspond to the coefficients $({{\alpha}_1}_{n;n{'}}^{\pm1;\pm2},{{\beta}_1}_{n;n{'}}^{\pm1;\pm2})$ and $({{\alpha}_3}_{n;n{'}}^{\pm1;\pm2},{{\beta}_3}_{n;n{'}}^{\pm1;\pm2})$ in \eqref{Zn1Zn'pm2}.}
\label{2ndcoefficients}
	\end{center}
\end{table*}

We then compared the contributions of each coupling term to the far-field WFE, as shown in Fig. \ref{AberrationCoupling}. 
\begin{figure*}[htbp]
	\begin{center}
		\includegraphics[width=0.7\textwidth]{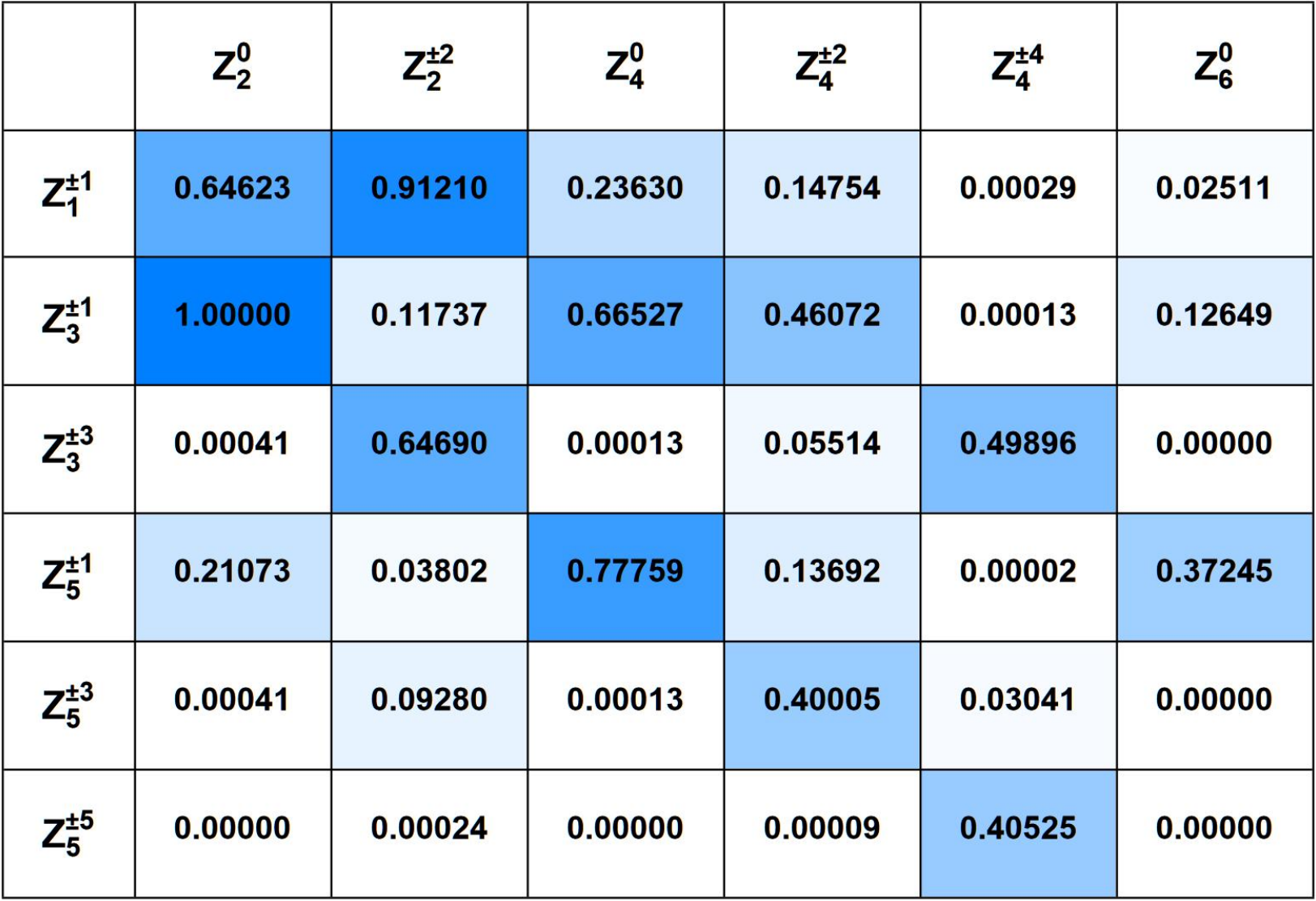}
	\end{center}
	\caption{The contribution of each coupling term ${Z^{2\alpha+1}_{\gamma}}{Z^{2\beta}_{{\gamma}{'}}}$ to the far-field WFE. Coefficients $a^{2\alpha+1}_{\gamma}$ and $a^{2\beta}_{{\gamma}{'}}$ are constrained to correspond to $\lambda/20$, meaning that all coefficients are equal to 0.157079 except for $a_4^0=0.20944$. }
\label{AberrationCoupling} 
\end{figure*}
From the figure, we draw the following conclusions:
\begin{enumerate}
\item The coupling between different terms varies, and not all coupling terms produce significant contributions, which is quite understandable. For ${Z^{2\alpha+1}_{\gamma}}{Z^{2\beta}_{{\gamma}{'}}}$, when $\left|2\alpha+1\right|-\left|2\beta\right|=3$, the lowest-order term in the Bessel function expansion is  $J_4(v)$. Its contribution is very small and acts only as a correction term among other coupling terms. Additionally, we can disregard low coupling terms that contribute minimally to the WFE.

\item For ${Z^{2\alpha+1}_{\gamma}}{Z^{2\beta}_{{\gamma}{'}}}$ with the same $\alpha$ and $\beta$, it is often observed that terms with smaller ${\gamma}$ and ${\gamma}{'}$ contribute more to the far-field WFE than those with larger ${\gamma}$ and ${\gamma}{'}$. Therefore, as the order of Zernike increases, the contribution of aberrations to the far-field WFE generally shows a decreasing trend; however, this statement does not strictly apply to the coupling terms. Therefore, the contributions of each coupling term still require specific analysis.

\item $Z_2^{\pm2}$, $Z_3^{\pm3}$, and $Z_4^{\pm4}$ are easily correctable on the ground, while axisymmetric terms are not; however, they may also arise due to thermal stress during in-orbit operations\cite{bely2003design}. Since $Z_1^{\pm1}$ usually serves as a correction for the optical axis, we can neglect it in this context. Given that higher-order terms are generally more stable than lower-order ones, we focus on the lower-order aberrations where $n\leq14$. Among these aberrations, $Z_2^{\pm2}$, $Z_3^{\pm3}$, and $Z_4^{\pm4}$ exhibit significant coupling primarily with themselves and relatively weak coupling with other aberrations. If telescopes are manufactured using materials with low thermal expansion coefficients, their impact on the far-field WFE should remain at a relatively low level. Conversely, the presence of $Z_3^{\pm1}$ and axisymmetric aberrations will be the primary contributors to the far-field WFE.
\end{enumerate}
\end{itemize}

\noindent For ${Z^{2{\alpha}_1+1}_{{\gamma}_1}}{Z^{2{\alpha}_2+1}_{{\gamma}_2}}$, we only consider $Z_1^{\pm1}Z_1^{\pm1}$, $Z_1^{\pm1}Z_3^{\pm1}$, $Z_1^{\pm1}Z_5^{\pm1}$, and $Z_3^{\pm1}Z_3^{\pm1}$. And we obtain:
\begin{equation}\label{Zn1Zn'1}
\begin{aligned}
	&\begin{pmatrix}
			Z_n^{1}Z_{n{'}}^{1} \\
			Z_n^{-1}Z_{n{'}}^{-1} \\
		\end{pmatrix}
		(r, \psi, z)=e^{-\frac{1}{2}}{\pi}^{\frac{1}{2}}(2\pi)\\
		&\left\{-\frac{1}{2}\left[{{\alpha}_1}_{n;n{'}}^{\pm1;\pm1}\frac{J_1(v)}{v}+{{\beta}_1}_{n;n{'}}^{\pm1;\pm1}\frac{J_3(v)}{v}\right]+\frac{1}{2}
	\begin{pmatrix}
			\cos2\psi \\
			-\cos2\psi \\
		\end{pmatrix}
	\left[{{\alpha}_2}_{n;n{'}}^{\pm1;\pm1}\frac{J_1(v)}{v}+{{\beta}_2}_{n;n{'}}^{\pm1;\pm1}\frac{J_3(v)}{v}\right]\right\},
\end{aligned}
\end{equation}
where
\begin{gather*}\label{co2ndOddOdd}
\left({\sigma}_{1;1}^{\pm1;\pm1},\;{\tau}_{1;1}^{\pm1;\pm1}\right)_1=a_1^{\pm1}a_1^{\pm1}\left(\num{0.122898},\;\num{-0.079482}\right),\left({\sigma}_{1;1}^{\pm1;\pm1},\;{\tau}_{1;1}^{\pm1;\pm1}\right)_2=a_1^{\pm1}a_1^{\pm1}\left(0.,\;\num{0.224087}\right),\\
\left({\sigma}_{1;3}^{\pm1;\pm1},\;{\tau}_{1;3}^{\pm1;\pm1}\right)_1=a_1^{\pm1}a_3^{\pm1}\left(\num{-0.0434156},\;\num{-0.245727}\right),\left({\sigma}_{1;3}^{\pm1;\pm1},\;{\tau}_{1;3}^{\pm1;\pm1}\right)_2=a_1^{\pm1}a_3^{\pm1}\left(0.,\;\num{0.0577399}\right),\\
\left({\sigma}_{1;5}^{\pm1;\pm1},\;{\tau}_{1;5}^{\pm1;\pm1}\right)_1=a_1^{\pm1}a_5^{\pm1}\left(\num{0.00478612},\;\num{0.0516885}\right),\left({\sigma}_{1;5}^{\pm1;\pm1},\;{\tau}_{1;5}^{\pm1;\pm1}\right)_2=a_1^{\pm1}a_5^{\pm1}\left(0.,\;\num{-0.0186651}\right),\\
\left({\sigma}_{3;3}^{\pm1;\pm1},\;{\tau}_{3;3}^{\pm1;\pm1}\right)_1=a_3^{\pm1}a_3^{\pm1}\left(\num{0.072286},\;\num{0.0326643}\right),\left({\sigma}_{3;3}^{\pm1;\pm1},\;{\tau}_{3;3}^{\pm1;\pm1}\right)_2=a_3^{\pm1}a_3^{\pm1}\left(0.,\;\num{0.092096}\right).\\
\end{gather*}

\begin{figure}
     \centering
     \begin{subfigure}[b]{0.3\textwidth}
         \centering
         \caption{$Z_1^{\pm1}Z_1^{\pm1}$}
         \includegraphics[width=1\textwidth]{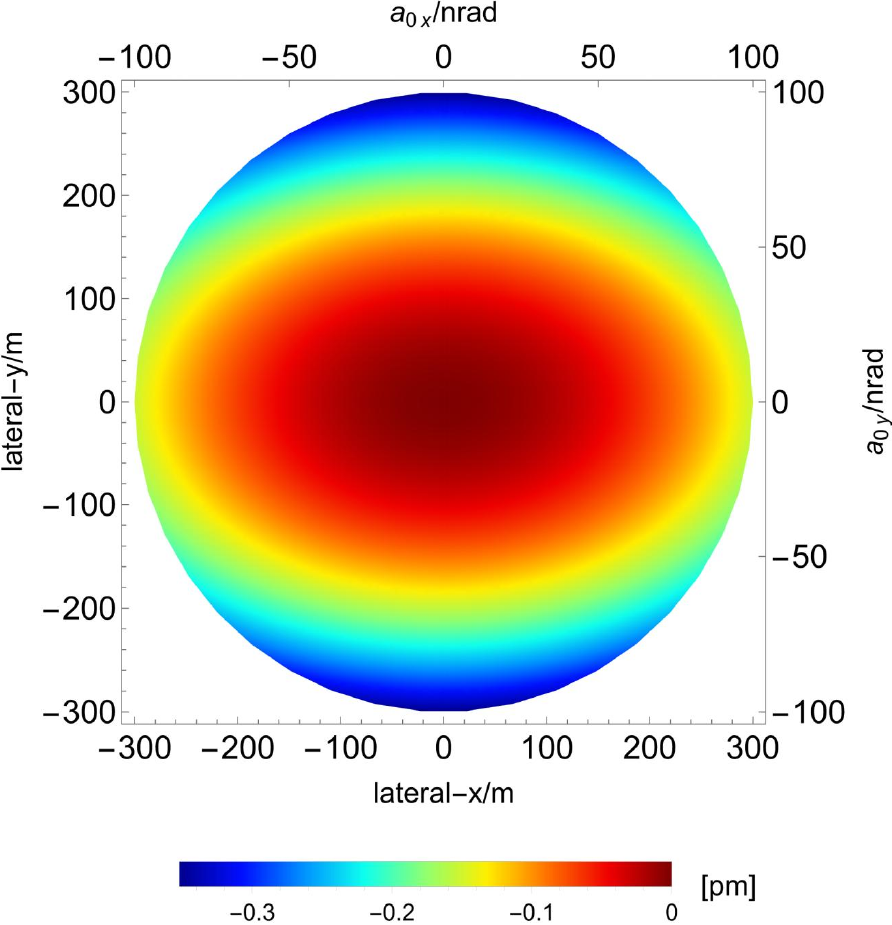}
         \label{fig:Z11Z11}
     \end{subfigure}
     \begin{subfigure}[b]{0.3\textwidth}
         \centering
         \caption{$Z_1^{\pm1}Z_3^{\pm1}$}
         \includegraphics[width=1\textwidth]{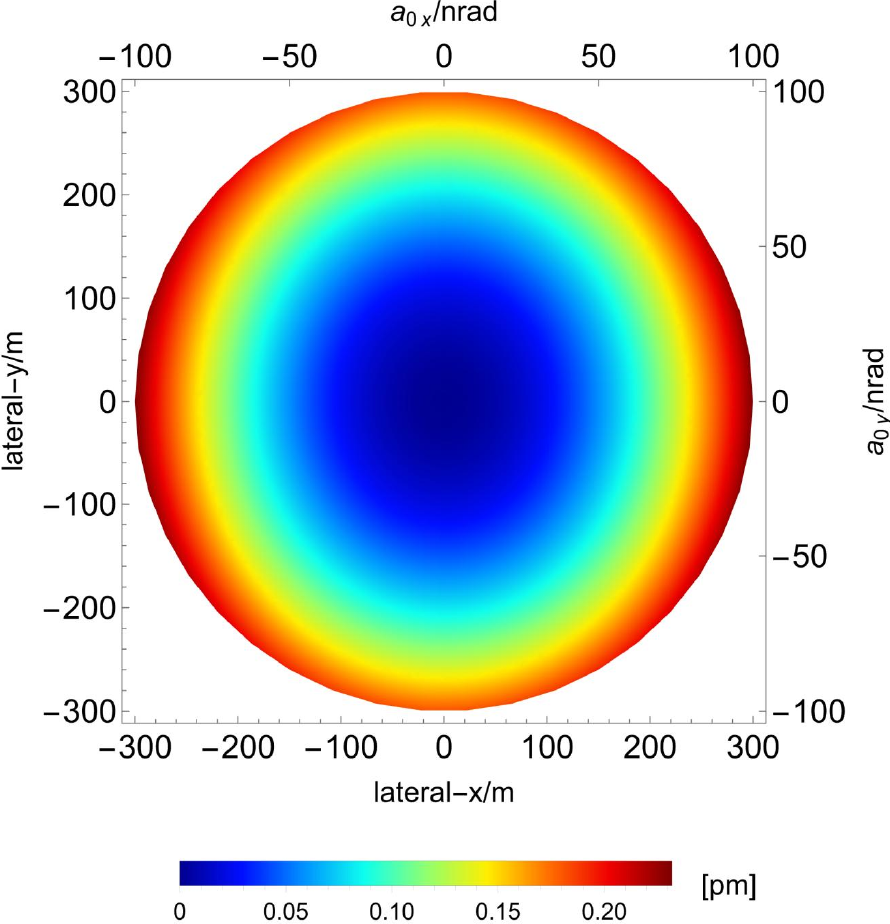}
         \label{fig:Z11Z31}
     \end{subfigure}
     
          \begin{subfigure}[b]{0.3\textwidth}
         \centering
         \caption{$Z_1^{\pm1}Z_5^{\pm1}$}
         \includegraphics[width=1\textwidth]{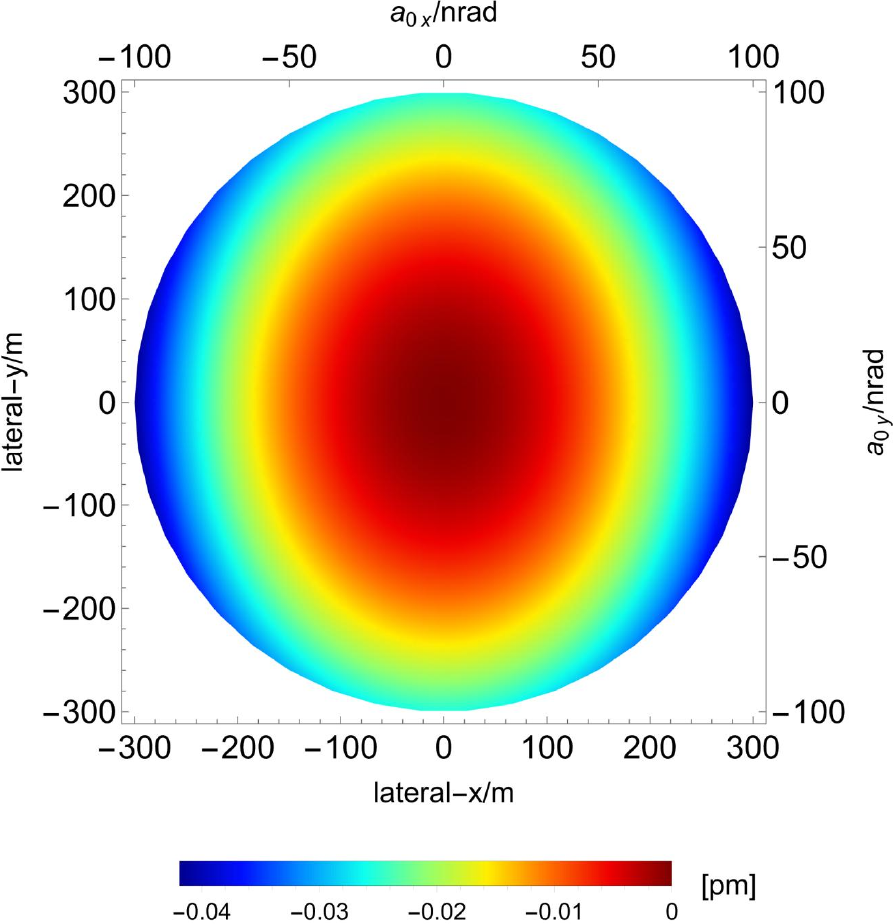}
         \label{fig:Z11Z51}        
     \end{subfigure}
          \begin{subfigure}[b]{0.3\textwidth}
         \centering
         \caption{$Z_3^{\pm1}Z_3^{\pm1}$}
         \includegraphics[width=1\textwidth]{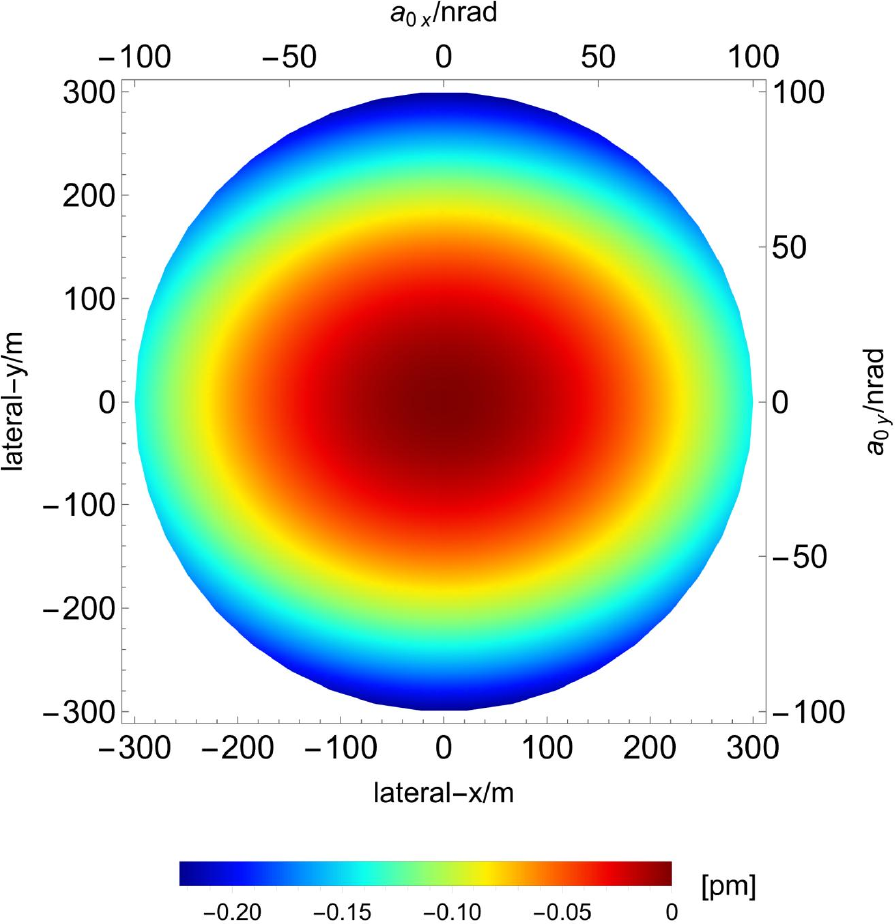}
         \label{fig:Z31Z31}        
     \end{subfigure}
        \caption{The Difference between $"{WFE}_{{Z^{2{\alpha}_1+1}_{{\gamma}_1}}{Z^{2{\alpha}_2+1}_{{\gamma}_2}}+Z_2^0}"$ and ${WFE}_{Z_2^0}$, calculated using A.E.. $a^0_2$ and all $a^{2\alpha+1}_{\gamma}$ are set to be 0.314159. The terms of $Z_{{\gamma}_1}^{\pm1}Z_{{\gamma}_2}^{\pm1}$only display $Z_{{\gamma}_1}^{1}Z_{{\gamma}_2}^{1}$. }
        \label{2ndOddOdd}
\end{figure}

As what we do for ${Z^{2\alpha+1}_{\gamma}}$, the difference between the "WFE" of ${Z^{2{\alpha}_1+1}_{{\gamma}_1}}{Z^{2{\alpha}_2+1}_{{\gamma}_2}}$ and the WFE of $Z_2^0$, as shown in Fig. \ref{2ndOddOdd}.
We can thus conclude that $Z_1^{\pm1}Z_5^{\pm1}$ can be discarded, while $Z_1^{\pm1}Z_1^{\pm1}$, $Z_1^{\pm1}Z_3^{\pm1}$, and $Z_3^{\pm1}Z_3^{\pm1}$ should be retained.

\subsection{Amplitude and Beam direction} \label{sbse:4.3}

It is reasonable to define the line connecting the beam spot center in the image plane and the beam origin as the optical axis. The beam propagates along this optical axis. Since the distortion of the  transmitted Gaussian beam is minimal, the spot of the distorted Gaussian beam in the far field can still be approximated as a Gaussian beam spot. Therefore, we consider the location of maximum amplitude to be the center of the beam spot. 

\begin{figure}
     \centering
     \begin{subfigure}[b]{0.3\textwidth}
         \centering
         \caption{$Z_1^{\pm1}$}
         \includegraphics[width=1\textwidth]{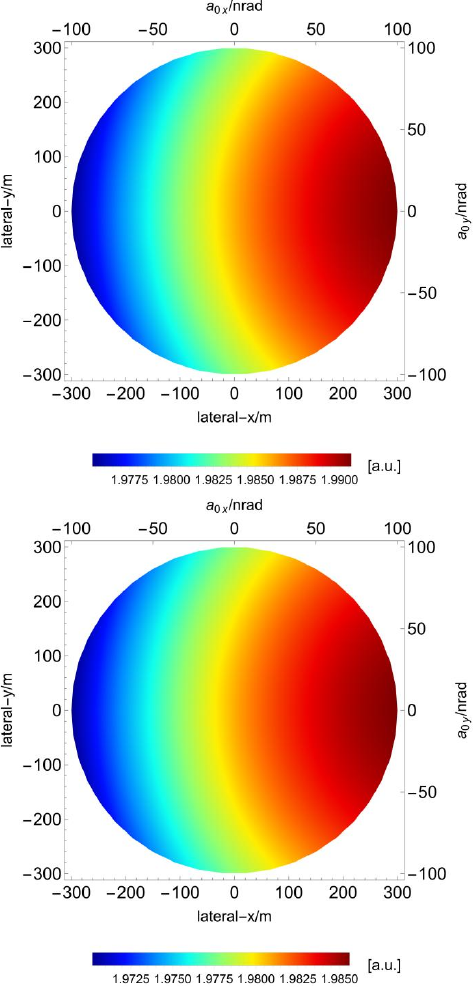}
         \label{fig:Z11amp}
     \end{subfigure}
     \begin{subfigure}[b]{0.3\textwidth}
         \centering
         \caption{$Z_3^{\pm1}$}
         \includegraphics[width=1\textwidth]{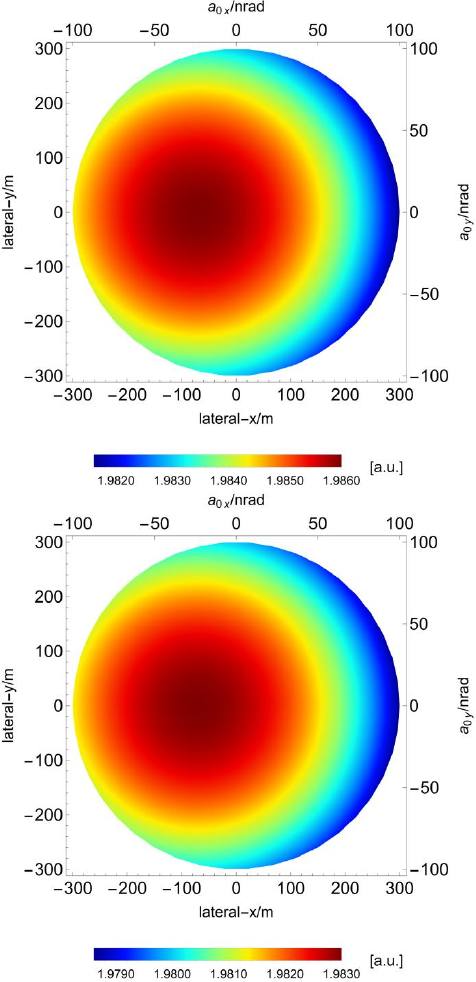}
         \label{fig:Z31amp}
     \end{subfigure}    
          \begin{subfigure}[b]{0.3\textwidth}
         \centering
         \caption{$Z_5^{\pm1}$}
         \includegraphics[width=1\textwidth]{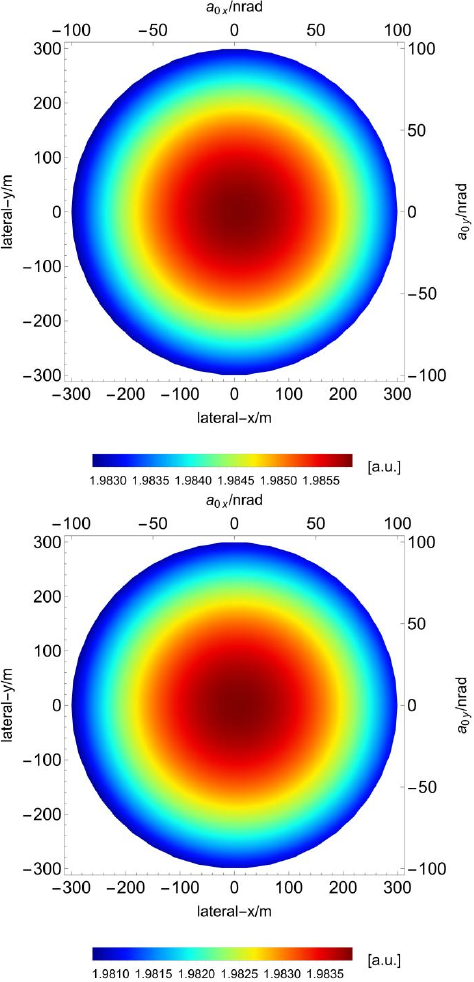}
         \label{fig:Z51amp}        
     \end{subfigure}
        \caption{The real part of "$U_0(r, \psi, z)+Z_n^1(r, \psi, z)$", where all $a^{\pm1}_{n}$ are set to 0.157079. The upper half of each subfigure is calculated using A.E. while the lower half is derived from N.I.. The terms of $Z_{{\gamma}}^{\pm1}$only display $Z_{{\gamma}}^{1}$. }
        \label{Zn1Amp}
\end{figure}

From the discussion in Subsection \ref{sbse:4.2}, we see that $Z_n^1$ introduces primary  non-spherical symmetric terms in the real part, causing the beam's amplitude to deviate from the coordinate center at the receiving side, as shown in Fig. \ref{Zn1Amp}.

We only consider $Z_n^{\pm1}$, $Z_1^{\pm1}Z_1^{\pm1}$, $Z_1^{\pm1}Z_3^{\pm1}$, $Z_3^{\pm1}Z_3^{\pm1}$, and neglect the terms containing $\cos2\psi$. By performing a Taylor expansion of $J_n(v)/v$ and retaining terms up to $v^2$, we obtain:
\begin{equation}\label{RealU}
Re\{U(r, \psi, z)\}=-av^2+bv\cos\psi+cv\sin\psi+d,
\end{equation}
where $v=\frac{k}{{z}}{r_a}r$, and

\begin{gather*}\label{coRealU}
a=\num{0.0307244}-\num{0.00466848}{(a_1^1)}^2-\num{0.00466848}{(a_1^{-1})}^2-\num{0.00240583}{a_1^1}{a_3^1}-\num{0.00240583}{a_1^{-1}}{a_3^{-1}},\\
b=\num{0.0614487}{a_1^1}- \num{0.0108539}{a_3^1} + \num{0.00119653}{a_5^1},\\
c=\num{0.0614487}{a_1^{-1}}- \num{0.0108539}{a_3^{-1}}+ \num{0.00119653}{a_5^{-1}},\\
d=\num{0.293997}-\num{0.0307244}{(a_1^1)}^2-\num{0.0307244}{(a_1^{-1})}^2 + \num{0.0217078}{a_1^1}{a_3^1}+\num{0.0217078}{a_1^{-1}}{a_3^{-1}}.\\
\end{gather*}
Thus, corresponding to the location of maximum amplitude, the coordinates of the beam center are: $(\frac{bz}{2akr_a},\frac{cz}{2akr_a})$.

We can add a new pair of X-Tilt and Y-Tilt to $a_1^{\pm1}$ in the transmitted WFE as compensation to shift the beam back to the point$(0,0)$, since for the required compensation shift, the following linear relationship between the pointing angle $\alpha$ and $a_1^{\pm1}$ holds:
\begin{equation}\label{tilt-a}
\alpha=\frac{a_1^{\pm1}}{kr_a}.
\end{equation}
Therefore, the new ${a{'}}_1^{1}$ and ${a{'}}_1^{-1}$ are:
\begin{equation}\label{compensationa11}
({a{'}}_1^{1},{a{'}}_1^{-1})=(a_1^1-\frac{b}{2a},a_1^{-1}-\frac{c}{2a}).
\end{equation}

\begin{figure*}[htbp]
	\begin{center}
		\includegraphics[width=0.9\textwidth]{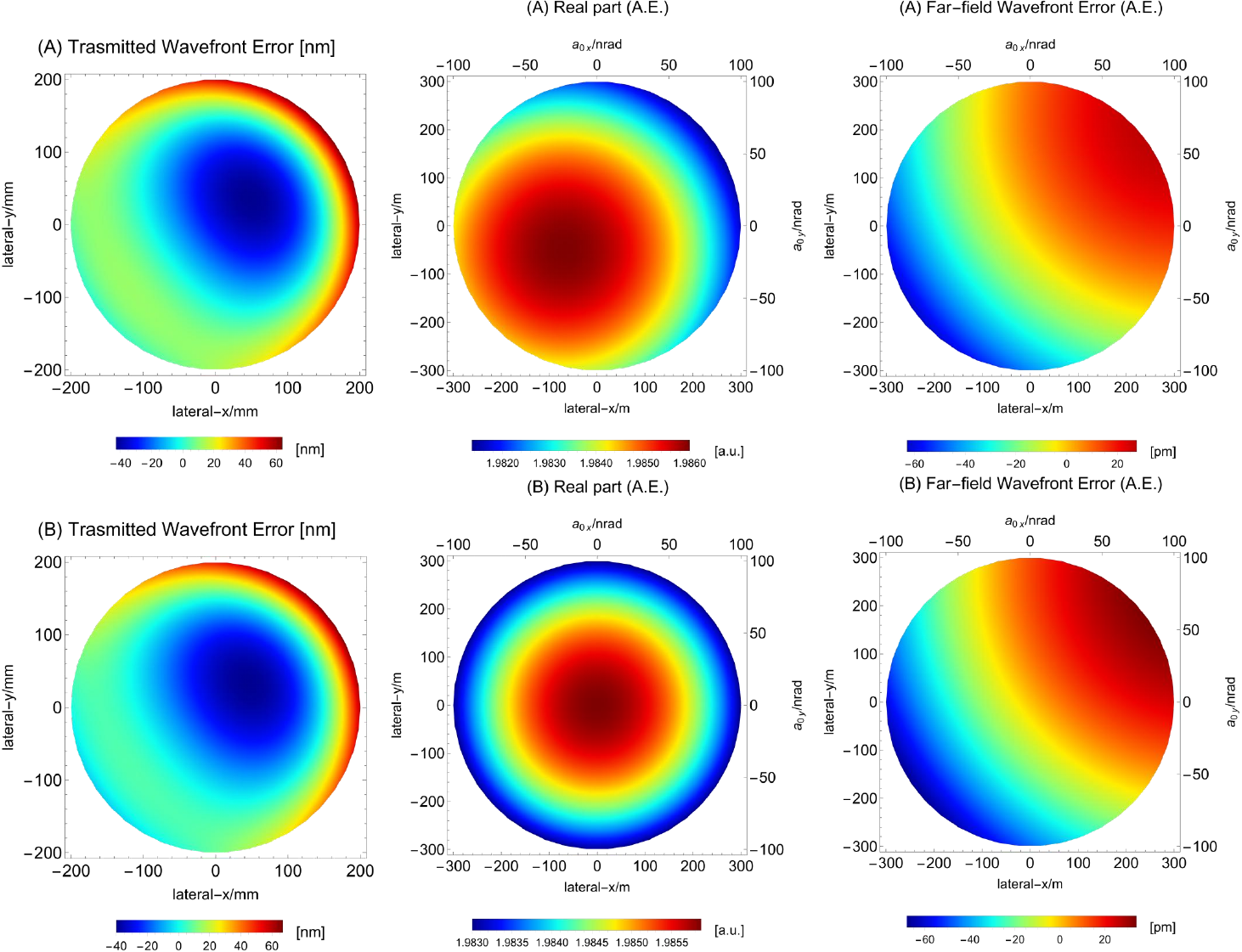}
	\end{center}
	\caption{The transmitted WFE, deviation of the optical axis, and far-field WFE of cases A and B. The transmitted WFE corresponds to $\lambda/10$ in both cases A and B. The far-field WFE (P-V) of cases A and B are $90.6068\;\text{pm}$ and $102.708\;\text{pm}$, respectively.}
	\label{Z31Ex} 
\end{figure*}

We then validate this result through an example involving double aberrations of $Z_3^{\pm1}$ and $Z_2^{0}$, while also addressing a question related to $Z_3^{\pm1}$. The coefficients corresponding to case A in Fig. \ref{Z31Ex} are listed in Table \ref{Z31ExT}. By performing calculations using \eqref{compensationa11}, the resulting deviation of the optical axis is effectively compensated, as shown in case B of Fig. \ref{Z31Ex}, where the coefficients of case B have been rescaled to $\lambda/10$.
\begin{table*}
\sisetup{round-mode=places, round-precision=5}
\abovetopsep=0pt
\aboverulesep=0pt
\belowrulesep=0pt
\belowbottomsep=0pt
\renewcommand\arraystretch{1.5}
	\begin{center}
			\setlength{\tabcolsep}{1mm}
		\begin{tabular}[width=0.7\textwidth]{c|c c c c c}
			\toprule[1.5pt]
			$Z^m_n$ & $Z_2^{0}$ &  $Z_1^1$ &  $Z_1^{-1}$ &  $Z_3^1$ &  $Z_3^{-1}$ \\ 
			\midrule[1.5pt]
			$a^m_n$ & $\num{0.188175}$ & $0.$ & $0.$ & $\num{0.152237}$& $\num{0.110607}$ \\
			\midrule[1.5pt]
			\midrule[1.5pt]
			$Z^m_n$ & $Z_2^{0}$ &  $Z_1^1$ &  $Z_1^{-1}$ &  $Z_3^1$ &  $Z_3^{-1}$ \\ 
			\midrule[1.5pt]
			$a^m_n$ & $\num{0.181358}$ & $\num{0.025916}$ & $\num{0.0188291}$ & $\num{0.146722}$& $\num{0.1066}$ \\
			\bottomrule[1.5pt]
		\end{tabular}
		\caption{The coefficient list of cases A and B. The corresponding transmitted WFE, deviation of the optical axis, and far-field WFE are shown in Fig. \ref{Z31Ex}.}
\label{Z31ExT}
	\end{center}
\end{table*}

However, correcting the optical axis results in a larger far-field WFE. As shown in Fig. \ref{Z31Ex}, the pre-correction value is $90.6068\;\text{pm}$, whereas the post-correction value is $102.708\;\text{pm}$. This result suggests that while correcting the optical axis on the ground may appear to address the issue of beam tilt, the presence of $Z_3^{\pm1}$ still critically affects the level of far-field WFE. Using beam tilt to correct the optical axis, in fact, results in an even greater far-field WFE. Therefore, it is essential to suppress $Z_3^{\pm1}$ itself rather than merely compensating for it with beam tilt.

\section{Result} \label{se:5}
In summary of Subsection \ref{sbse:4.2}, we derive a formula for the far-field amplitude and WFE that incorporates the contributions of the first 21 Zernike polynomial aberrations:
\begin{subequations}\label{WFE2}
\begin{align}
&|E(r, \psi, z)|=Re\{E(r, \psi, z)\}={r_a}^2Re\left\{U(r, \psi, z)\right\},\\
&{\delta}{\Theta}(r, \psi, z)=\frac{\lambda}{2\pi}\left(\frac{Im\left\{U(r, \psi, z)\right\}}{Re\left\{U(r, \psi, z)\right\}}-\frac{Im\left\{U(0, \psi, z)\right\}}{Re\left\{U(0, \psi, z)\right\}}\right),
 \label{WFE2_1}
\\
\begin{split}
	Im\left\{U(r, \psi, z)\right\}=&A_0\frac{J_1(v)}{v}+(A_1{\cos\phi}+A_2{\sin\phi})\frac{J_2(v)}{v}+(A_3+A_4{\cos2\phi}+A_5{\sin2\phi})\frac{J_3(v)}{v}\\
	&+(A_6{\cos\phi}+A_7{\sin\phi}+A_8{\cos3\phi}+A_9{\sin3\phi})\frac{J_4(v)}{v},\\
\end{split}
\\
\begin{split}
	Re\left\{U(r, \psi, z)\right\}=&B_0\frac{J_1(v)}{v}+(B_1{\cos\phi}+B_2{\sin\phi})\frac{J_2(v)}{v}+(B_3+B_4{\cos2\phi})\frac{J_3(v)}{v}\\
	&+(B_5{\cos\phi}+B_6{\sin\phi}+B_7{\cos3\phi}+B_8{\sin3\phi})\frac{J_4(v)}{v}.\\
\end{split}
\end{align}
\end{subequations}
The coefficients $A_i$ and $B_i$ are provided in Appendix \ref{se:7}.

\begin{figure*}[htbp]
	\begin{center}
		\includegraphics[width=0.9\textwidth]{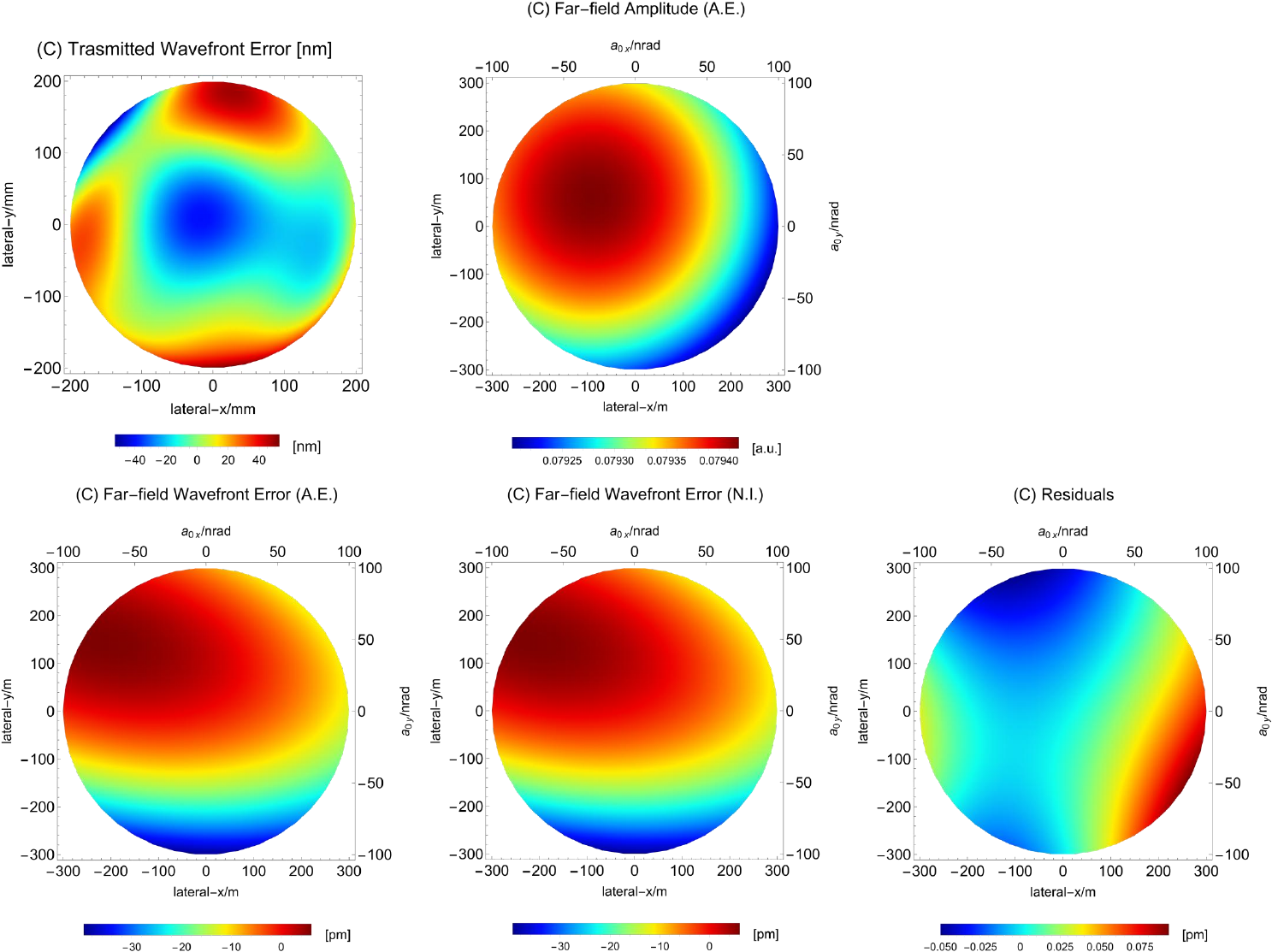}
	\end{center}
	\caption{The transmitted WFE, far-field amplitude, far-field WFE (A.E), far-field WFE (N.I), and the far-field WFE residuals between A.E. and N.I.. The optical axis is not compensated. The coefficients are listed in Table \ref{Exapledcoefficients1}.}
	\label{Example1} 
\end{figure*}

We illustrate the coupling noise model discussion through the following example. In this example, the randomly generated transmitted WFE is constrained to $\lambda/10$ (P V). The coefficients for each Zernike aberration are listed in Table \ref{Exapledcoefficients1}. The transmitted WFE, as well as its far-field amplitude, far-field WFE, and the comparison of far-field WFEs calculated using A.E. and N.I., are all presented in Fig. \ref{Example1}.
\begin{table*}
\sisetup{round-mode=places, round-precision=5}
\abovetopsep=0pt
\aboverulesep=0pt
\belowrulesep=0pt
\belowbottomsep=0pt
\renewcommand\arraystretch{1.5}
	\begin{center}
		\setlength{\tabcolsep}{1mm}
		\begin{tabular}{c|c c c c c c c}
			\toprule[1.5pt]
			${Z^m_n}$ &  ${Z_1^1}$ &  ${Z_1^{-1}}$ & ${Z_2^{0}}$ &  ${Z_2^2}$ &  ${Z_2^{-2}}$ &  ${Z_3^1}$ &  ${Z_3^{-1}}$\\ 
			\midrule[1.5pt]
			${a^m_n}$ & $\num{-0.0290403}$ & $\num{0.0135543}$ & $\num{0.145519}$ & $\num{-0.105273}$& $\num{0.0566279}$ & $\num{0.0158143}$ & $\num{-0.048233}$ \\
			\midrule[1.5pt]
			 &  ${Z_3^3}$ &  ${Z_3^{-3}}$ & ${Z_4^{0}}$ &  ${Z_4^2}$ &  ${Z_4^{-2}}$ &  ${Z_4^4}$ &  ${Z_4^{-4}}$ \\ 
			\midrule[1.5pt]
			 & $\num{-0.112372}$ & $\num{-0.0447792}$ & $\num{-0.0677458}$ & $\num{0.0158143}$& $\num{0.0124993}$ & $\num{0.0975962}$ & $\num{0.00259626}$ \\
			\midrule[1.5pt]
			 &  ${Z_5^1}$ &  ${Z_5^{-1}}$ & ${Z_5^{3}}$ &  ${Z_5^{-3}}$ &  ${Z_5^{5}}$ &  ${Z_5^{-5}}$ &  ${Z_6^{0}}$\\ 
			\midrule[1.5pt]
			 & $\num{0.0689143}$ & $\num{-0.0733618}$ & $\num{0.0318497}$ & $\num{-0.0299652}$& $\num{-0.0109276}$ & $\num{0.00481528}$ & $\num{0.0158143}$ \\
			\bottomrule[1.5pt]
		\end{tabular}
		\caption{The coefficients for each Zernike aberration. The corresponding transmitted WFE is shown in Fig. \ref{Example1}.}
\label{Exapledcoefficients1}
	\end{center}
\end{table*}

\begin{figure*}[htbp]
	\begin{center}
		\includegraphics[width=0.9\textwidth]{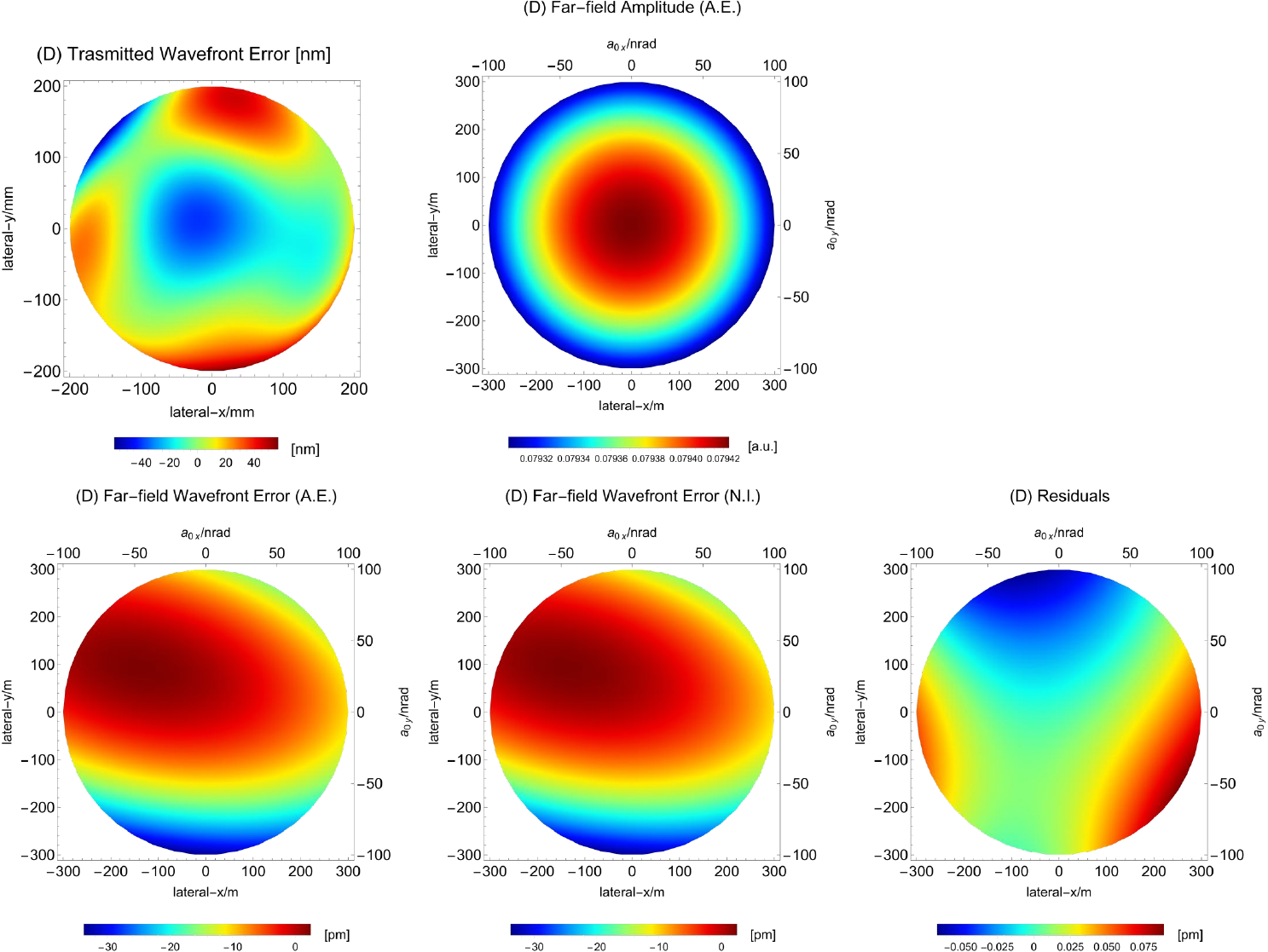}
	\end{center}
	\caption{The updated transmitted WFE, far-field amplitude, far-field WFE (A.E), far-field WFE (N.I), and the far-field WFE residuals between A.E. and N.I., whose optical axis is compensated. The updated values of $a_1^{1}$ and $a_1^{-1}$ are 0.00145 and -0.00709, respectively.}
	\label{Example2} 
\end{figure*}

After correcting the optical axis direction using \eqref{compensationa11}, we obtain the updated coefficients of $a_1^{\pm1}$, with $\boldsymbol{a_1^{1}=0.00145}$ and $\boldsymbol{a_1^{-1}=-0.00709}$. The corrected far-field amplitude and WFE are shown in Fig.  \ref{Example2}.

\begin{figure*}[htbp]
	\begin{center}
		\includegraphics[width=0.9\textwidth]{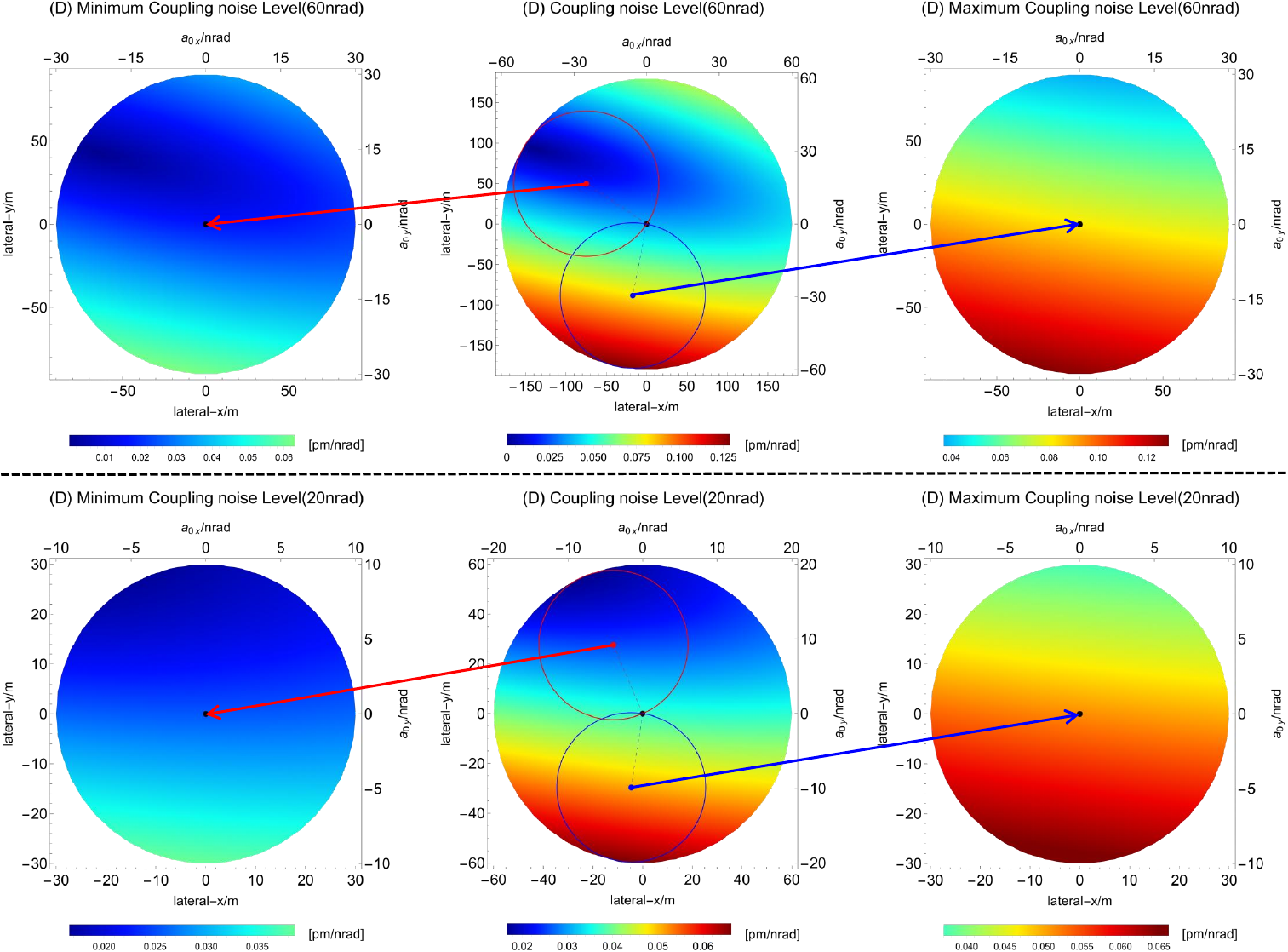}
	\end{center}
	\caption{The far-field 	WFE corresponds to the results in Fig. \ref{Example2}. The middle figure show the noise levels of coupling noise within a pointing angle range of 60 nrad (up) and 20 nrad (down), respectively. The pointing angles of 60 nrad and 20 nrad are composed of a 30 nrad static pointing angle plus 30 nrad laser jitter, and a 10 nrad static pointing angle plus 10 nrad laser jitter, respectively. The figures on the left and right sides respectively correspond to the minimum and maximum noise levels within the laser jitter ranges of 30 nrad (up) and 10 nrad (down).}
	\label{Ex2Noiselevel} 
\end{figure*}

Next, we examine two cases where both the static pointing angle and laser jitter are at levels of $10\;nrad/\sqrt{Hz}$ and 30 $30\;nrad/\sqrt{Hz}$. To facilitate the discussion, we approximate \eqref{WFE2_1} by performing a Taylor expansion of $J_n(v)/v$ and retaining terms up to $v^2$. In the Cartesian coordinate system, we obtain:
\begin{equation}\label{WFE3}
\begin{split}
&{\delta}{\Theta}(x,y,z)\\
&=\frac{\lambda}{2\pi}\left(\frac{A_0(\frac{1}{2}-\frac{v^2}{16})+(A_1{\cos\phi}+A_2{\sin\phi})\frac{v}{8}+(A_3+A_4{\cos2\phi}+A_5{\sin2\phi})\frac{v^2}{48}}{B_0(\frac{1}{2}-\frac{v^2}{16})+(B_1{\cos\phi}+B_2{\sin\phi})\frac{v}{8}+(B_3+B_4{\cos2\phi})\frac{v^2}{48}}-\frac{A_0}{B_0}\right)\\
&=\frac{\lambda}{2\pi}\left(\frac{24A_0+6A_1v_x+6A_2v_y+(A_3-3A_0+A_4){v_x}^2+(A_3-3A_0-A_4){v_y}^2+2A_5v_xv_y}{24B_0+6B_1v_x+6B_2v_y+(B_3-3B_0+B_4){v_x}^2+(B_3-3B_0-B_4){v_y}^2}-\frac{A_0}{B_0}\right),
\end{split}
\end{equation}
where $v_x=\frac{k}{{z}}{r_a}x$ and $v_y=\frac{k}{{z}}{r_a}y$. By applying \eqref{NoiseLevel}, we obtain the noise levels of the coupling noise within a displacement range of $180\;\text{m}\;(60\;\text{nrad})$ and $60\;\text{m}\;(20\;\text{nrad})$. Fig. \ref{Ex2Noiselevel} illustrates the maximum and minimum noise levels in these two cases.

Thus, the above process illustrates the establishment of a transfer function that relates the initial parameters—transmitted WFE, static pointing angle, and laser jitter—to both the far-field WFE and the coupling noise from laser pointing jitter. The model presented in this paper provides a comprehensive framework for discussion.

\section{Conclusions and Summary} \label{se:6}
In this paper, we analytically derive an approximate expression for the far-field diffraction integral of distorted Gaussian beams based on the Nijboer-Zernike theory, accounting for the first 21 orders of Zernike aberrations. Within a far-field WFE range of $100\;\text{nrad}$, this approximate expression achieves an error level of approximately $0.1\;\text{pm}$ compared to numerical integration, which is sufficient to meet the requirements. Additionally, as an analytical formula, it demonstrates rapid computational response capabilities, making it a useful tool for simulating entire laser link systems, including TTL noise link analysis.

The approximate expression offers valuable physical insights into how different Zernike aberrations influence far-field WFE. Through a term-by-term analysis, we find that a second-order expansion of the distortion terms suffices to meet precision requirements. Building on this, we calculate the coupling relationships between different orders of aberrations and determine the contribution coefficients of each coupling term to the far-field WFE. First, we identify that only certain couplings need to be considered based on the varying contribution coefficients and we provide theoretical explanations for this finding. Second, we observe that the contribution coefficients do not decrease with increasing order of the Zernike terms. Third, we propose that among the lower-order aberrations $(n\leq14)$, $Z_2^{\pm2}$, $Z_3^{\pm3}$, and $Z_4^{\pm4}$ primarily couple with each other while exhibiting weaker coupling with other aberrations. When telescopes are constructed from materials with low thermal expansion coefficients, their impact on the far-field WFE can be mitigated. In this scenario, $Z_3^{\pm1}$ and axisymmetric aberrations, along with their coupling, emerge as primary contributors. Subsequently, in our analysis of optical axis correction, we find that correcting the optical axis offset caused by $Z_3^{\pm1}$ through beam tilt results in an increase in the far-field WFE. Therefore, we conclude that the presence of $Z_3^{\pm1}$ significantly influences the level of far-field WFE and needs to be actively suppressed rather than simply compensated by beam tilt. These conclusions provide valuable guidance for the design and assembly of telescope.

We have established a noise model for far-field WFE and laser pointing jitter coupling noise, encompassing three main influencing factors: transmitted WFE, static pointing angle, and laser jitter. The model utilizes the derived approximate expression to relate the Zernike coefficients of the transmitted WFE to the far-field WFE, resulting in an approximate noise formula. This result can be applied not only to the link model of TTL noise, but also facilitates the establishment of a parameter space that connects the three factors, providing theoretical support for future system optimization.
\section{Acknowledgements}
This work has been supported in part by the National Key Research and Development Program of China under Grant No.2020YFC2201501, the National Science Foundation of China (NSFC) under Grants No. 12147103 (special fund to the center for quanta-to-cosmos theoretical physics), No. 11821505, the Strategic Priority Research Program of the Chinese Academy of Sciences under Grant No. XDB23030100, and the Chinese Academy of Sciences (CAS). 
\appendix

\section{$A_i$ and $B_i$ coefficients} \label{se:7}
\begin{subequations}\label{WFE2CoAi}
\sisetup{round-mode=places, round-precision=5}
\begin{align}
\begin{split}
A_0=\num{-0.0964035} a_2^0 + \num{0.00964035} {(a_2^0)}^3 + \num{0.00957224} a_4^0,\\
\end{split}
\\
\begin{split}
&A_1=\\
&\num{0.105976} a_1^1 a_2^0 + \num{0.149391} a_1^1 a_2^2 + \num{0.149391} a_1^{-1} a_2^{-2} + \num{0.163818} a_2^0 a_3^1 + \num{0.0192466} a_2^2 a_3^1 + \\
 &\num{0.10601} a_2^2 a_3^3 + \num{0.0192466} a_2^{-2} a_3^{-1} + \num{0.10601} a_2^{-2} a_3^{-3} - \num{0.0289891} a_1^1 a_4^0 + \num{0.0817722} a_3^1 a_4^0 + \\
 &\num{-0.0241349} a_1^1 a_4^2 + \num{0.0754478} a_3^1 a_4^2 + \num{0.00905931} a_3^3 a_4^2 + \num{0.0817722} a_3^3 a_4^4 - \num{0.0241349} a_1^{-1} a_4^{-2} + \\
 &\num{0.0754478} a_3^{-1} a_4^{-2} + \num{0.00905931} a_3^{-3} a_4^{-2} + \num{0.0817722} a_3^{-3} a_4^{-4} -\num{0.034459} a_2^0 a_5^1 - \num{0.00622169} a_2^2 a_5^1 + \\
 &\num{0.0955441} a_4^0 a_5^1 + \num{0.0224356} a_4^2 a_5^1 - \num{0.0151783} a_2^2 a_5^3 + \num{0.0655388} a_4^2 a_5^3 + \num{0.00500097} a_4^4 a_5^3 + \\
  &\num{0.0664179} a_4^4 a_5^5 - \num{0.00622169} a_2^{-2} a_5^{-1} + \num{0.0224356} a_4^{-2} a_5^{-1} - \num{0.0151783} a_2^{-2} a_5^{-3} + \num{0.0655388} a_4^{-2} a_5^{-3} + \\
 &\num{0.00500097} a_4^{-4} a_5^{-3} +  \num{0.0664179} a_4^{-4} a_5^{-5} + \num{0.00410239} a_1^1 a_6^0 - \num{0.0206871} a_3^1 a_6^0 + \num{0.0610328} a_5^1 a_6^0,\\
\end{split}
\\
\begin{split}
&A_2=\\
&\num{0.105976} a_1^{-1} a_2^0 - \num{0.149391} a_1^{-1} a_2^2 + \num{0.149391} a_1^1 a_2^{-2} + \num{0.0192466} a_2^{-2} a_3^1 - \num{0.10601} a_2^{-2} a_3^3 +\\ 
 &\num{0.163818} a_2^0 a_3^{-1} - \num{0.0192466} a_2^2 a_3^{-1} + \num{0.10601} a_2^2 a_3^{-3} - \num{0.0289891} a_1^{-1} a_4^0 + \num{0.0817722} a_3^{-1} a_4^0 + \\
 &\num{0.0241349} a_1^{-1} a_4^2 - \num{0.0754478} a_3^{-1} a_4^2 + \num{0.00905931} a_3^{-3} a_4^2 - \num{0.0817722} a_3^{-3} a_4^4 - \num{0.0241349} a_1^1 a_4^{-2} + \\
 &\num{0.0754478} a_3^1 a_4^{-2} - \num{0.00905931} a_3^3 a_4^{-2} + \num{0.0817722} a_3^3 a_4^{-4} - \num{0.00622169} a_2^{-2} a_5^1 + \num{0.0224356} a_4^{-2} a_5^1 + \\
 &\num{0.0151783} a_2^{-2} a_5^3 - \num{0.0655388} a_4^{-2} a_5^3 + \num{0.00500097} a_4^{-4} a_5^3 - \num{0.0664179} a_4^{-4} a_5^5 - \num{0.034459} a_2^0 a_5^{-1} + \\
 &\num{0.00622169} a_2^2 a_5^{-1} + \num{0.0955441} a_4^0 a_5^{-1} - \num{0.0224356} a_4^2 a_5^{-1} - \num{0.0151783} a_2^2 a_5^{-3} + \num{0.0655388} a_4^2 a_5^{-3} - \\
 &\num{0.00500097} a_4^4 a_5^{-3} + \num{0.0664179} a_4^4 a_5^{-5} + \num{0.00410239} a_1^{-1} a_6^0 - \num{0.0206871} a_3^{-1} a_6^0 + \num{0.0610328} a_5^{-1} a_6^0,\\
\end{split}
\\
\begin{split}
&A_3=-\num{0.607138} a_2^0 + \num{0.0615342} {(a_2^0)}^3 + \num{0.115684} a40 - \num{0.0123072} a_6^0,\\
\end{split}
\\
\begin{split}
&A_4=-\num{0.448174} a_2^2+\num{0.0724048} a_4^2,\\
\end{split}
\\
\begin{split}
&A_5=-\num{0.448174} a_2^{-2}+\num{0.0724048} a_4^{-2},\\
\end{split}
\\
\begin{split}
&A_6=\\
&-\num{0.327636} a_1^1 a_2^0- \num{0.0384933} a_1^1 a_2^2- \num{0.0384933} a_1^{-1} a_2^{-2}+ \num{0.0871047} a_2^0 a_3^1- \num{0.122795} a_2^2 a_3^1- \\
 &\num{0.0665939} a_2^2 a_3^3- \num{0.122795} a_2^{-2} a_3^{-1}- \num{0.0665939} a_2^{-2} a_3^{-3}- \num{0.163544} a_1^1 a_4^0- \num{0.0775725} a_3^1 a_4^0- \\
 &\num{0.150896} a_1^1 a_4^2+ \num{0.0441091} a_3^1 a_4^2- \num{0.085894} a_3^3 a_4^2- \num{0.0714189} a_3^3 a_4^4- \num{0.150896} a_1^{-1} a_4^{-2}+ \\
 &\num{0.0441091} a_3^{-1} a_4^{-2}- \num{0.085894} a_3^{-3} a_4^{-2}- \num{0.0714189} a_3^{-3} a_4^{-4}- \num{0.235119} a_2^0 a_5^1- \num{0.0305425} a_2^2 a_5^1+ \\
 &\num{0.0270358} a_4^0 a_5^1- \num{0.0591804} a_4^2 a_5^1- \num{0.090719} a_2^2 a_5^3+ \num{0.0249651} a_4^2 a_5^3- \num{0.061798} a_4^4 a_5^3- \\
 &\num{0.0694947} a_4^4 a_5^5- \num{0.0305425} a_2^{-2} a_5^{-1}- \num{0.0591804} a_4^{-2} a_5^{-1}- \num{0.090719} a_2^{-2} a_5^{-3}+ \num{0.0249651} a_4^{-2} a_5^{-3}- \\
& \num{0.061798} a_4^{-4} a_5^{-3}- \num{0.0694947} a_4^{-4} a_5^{-5}+ \num{0.0413741} a_1^1 a_6^0- \num{0.130511} a_3^1 a_6^0- \num{0.0350193} a_5^1 a_6^0,\\
\end{split}
\\
\begin{split}
&A_7=\\
&-\num{0.327636} a_1^{-1} a_2^0 + \num{0.0384933} a_1^{-1} a_2^2 - \num{0.0384933} a_1^1 a_2^{-2} - \num{0.122795} a_2^{-2} a_3^1 + \num{0.0665939} a_2^{-2} a_3^3 + \\
& \num{0.0871047} a_2^0 a_3^{-1} + \num{0.122795} a_2^2 a_3^{-1} - \num{0.0665939} a_2^2 a_3^{-3} - \num{0.163544} a_1^{-1} a_4^0 - \num{0.0775725} a_3^{-1} a_4^0 + \\
& \num{0.150896} a_1^{-1} a_4^2 - \num{0.0441091} a_3^{-1} a_4^2 - \num{0.085894} a_3^{-3} a_4^2 + \num{0.0714189} a_3^{-3} a_4^4 - \num{0.150896} a_1^1 a_4^{-2} + \\
& \num{0.0441091} a_3^1 a_4^{-2} + \num{0.085894} a_3^3 a_4^{-2} - \num{0.0714189} a_3^3 a_4^{-4} - \num{0.0305425} a_2^{-2} a_5^1 - \num{0.0591804} a_4^{-2} a_5^1 + \\
& \num{0.090719} a_2^{-2} a_5^3 - \num{0.0249651} a_4^{-2} a_5^3 - \num{0.061798} a_4^{-4} a_5^3 + \num{0.0694947} a_4^{-4} a_5^5 - \num{0.235119} a_2^0 a_5^{-1} + \\
& \num{0.0305425} a_2^2 a_5^{-1} + \num{0.0270358} a_4^0 a_5^{-1} + \num{0.0591804} a_4^2 a_5^{-1} - \num{0.090719} a_2^2 a_5^{-3} + \num{0.0249651} a_4^2 a_5^{-3} + \\
& \num{0.061798} a_4^4 a_5^{-3} - \num{0.0694947} a_4^4 a_5^{-5} + \num{0.0413741} a_1^{-1} a_6^0 - \num{0.130511} a_3^{-1} a_6^0 - \num{0.0350193} a_5^{-1} a_6^0,\\
\end{split}
\\
\begin{split}
&A_8=\\
&-\num{0.21202} a_1^1 a_2^2 + \num{0.21202} a_1^{-1} a_2^{-2} - \num{0.0665939} a_2^2 a_3^1 + \num{0.0665939} a_2^{-2} a_3^{-1} - \num{0.0181186} a_1^1 a_4^2 - \\
& \num{0.085894} a_3^1 a_4^2 + \num{0.0181186} a_1^{-1} a_4^{-2} + \num{0.085894} a_3^{-1} a_4^{-2} - \num{0.00188526} a_2^2 a_5^1 - \num{0.0496409} a_4^2 a_5^1 + \\
& \num{0.00188526} a_2^{-2} a_5^{-1} + \num{0.0496409} a_4^{-2} a_5^{-1},\\
\end{split}
\\
\begin{split}
&A_9=\\
&-\num{0.21202} a_1^{-1} a_2^2  - \num{0.21202} a_1^1 a_2^{-2}  - \num{0.0665939} a_2^{-2} a_3^1  - \num{0.0665939} a_2^2 a_3^{-1}  - \num{0.0181186} a_1^{-1} a_4^2  - \\
& \num{0.085894} a_3^{-1} a_4^2  - \num{0.0181186} a_1^1 a_4^{-2}  - \num{0.085894} a_3^1 a_4^{-2}  - \num{0.00188526} a_2^{-2} a_5^1  - \num{0.0496409} a_4^{-2} a_5^1  - \\
& \num{0.00188526} a_2^2 a_5^{-1}  - \num{0.0496409} a_4^{-2} a_5^{-1}.\\
\end{split}
\end{align}
\end{subequations}
\begin{subequations}\label{WFE2CoBi}
\sisetup{round-mode=places, round-precision=5}
\begin{align}
\begin{split}
&B_0=\num{0.587993} - \num{0.0614487} {(a_1^1)}^2 - \num{0.0614487} {(a_1^{-1})}^2 - \num{0.10119} {(a_2^0)}^2 + 
 \num{0.0217078} a_1^1 a_3^1 + \\
& \num{0.0217078} a_1^{-1} a_3^{-1},\\
\end{split}
\\
\begin{split}
&B_1=\num{0.49159} a_1^1 - \num{0.0868312} a_3^1 + \num{0.00957224} a_5^1,\\
\end{split}
\\
\begin{split}
&B_2=\num{0.49159} a_1^{-1} - \num{0.0868312} a_3^{-1} + \num{0.00957224} a_5^{-1},\\
\end{split}
\\
\begin{split}
&B_3=\num{0.28921} + \num{0.0397409} ({a_1^1)}^2 + \num{0.0397409} ({a_1^{-1})}^2 - \num{0.0867631} ({a_2^0)}^2 + \num{0.122863} a_1^1 a_3^1 + \\
&\num{0.122863} a_1^{-1} a_3^{-1},\\
\end{split}
\\
\begin{split}
&B_4=\num{0.112043} ({a_1^1})^2 + \num{0.112043} ({a_1^{-1}})^2 + \num{0.02887} a_1^1 a_3^1 + \num{0.02887} a_1^{-1} a_3^{-1},\\
\end{split}
\\
\begin{split}
&B_5=\num{0.173662} a_1^1 - \num{0.578285} a_3^1 + \num{0.112949} a_5^1,\\
\end{split}
\\
\begin{split}
&B_6=\num{0.173662} a_1^{-1} - \num{0.578285} a_3^{-1} + \num{0.112949} a_5^{-1},\\
\end{split}
\\
\begin{split}
&B_7=\num{0.424039} a_3^3 ,\\
\end{split}
\\
\begin{split}
&B_8=\num{0.424039} a_3^{-3} ,\\
\end{split}
\end{align}
\end{subequations}

\bibliography{reference,library}
\makeatletter
\end{document}